\documentclass[twocolumn]{aastex631}
\submitjournal{AJ}
\shorttitle{Four Short-Period Sub-Neptunes Orbiting M dwarfs}
\shortauthors{Hori et al.}
\graphicspath{{./}{figures/}}

\begin{document}

\title{The Discovery and Follow-up of Four Transiting Short-period Sub-Neptunes Orbiting M dwarfs}

\correspondingauthor{Yasunori Hori}
\email{yasunori.hori@nao.ac.jp}

\author[0000-0003-4676-0251]{Yasunori Hori}
\affiliation{Astrobiology Center, 2-21-1 Osawa, Mitaka, Tokyo 181-8588, Japan}
\affiliation{National Astronomical Observatory of Japan, 2-21-1 Osawa, Mitaka, Tokyo 181-8588, Japan}
\affiliation{Department of Astronomical Science, The Graduated University for Advanced Studies (SOKENDAI), 2-21-1 Osawa, Mitaka, Tokyo 181-8588, Japan}

\author[0000-0002-4909-5763]{Akihiko Fukui}
\affiliation{Komaba Institute for Science, The University of Tokyo, 3-8-1 Komaba, Meguro, Tokyo 153-8902, Japan}
\affiliation{Instituto de Astrof\'{i}sica de Canarias (IAC), E-38200 La Laguna, Tenerife, Spain\label{IAC}}

\author[0000-0003-3618-7535]{Teruyuki Hirano}
\affiliation{Astrobiology Center, 2-21-1 Osawa, Mitaka, Tokyo 181-8588, Japan}
\affiliation{National Astronomical Observatory of Japan, 2-21-1 Osawa, Mitaka, Tokyo 181-8588, Japan}
\affiliation{Department of Astronomical Science, The Graduated University for Advanced Studies (SOKENDAI), 2-21-1 Osawa, Mitaka, Tokyo 181-8588, Japan}

\author[0000-0001-8511-2981]{Norio Narita}
\affiliation{Komaba Institute for Science, The University of Tokyo, 3-8-1 Komaba, Meguro, Tokyo 153-8902, Japan}
\affiliation{Astrobiology Center, 2-21-1 Osawa, Mitaka, Tokyo 181-8588, Japan}
\affiliation{Instituto de Astrof\'{i}sica de Canarias (IAC), E-38200 La Laguna, Tenerife, Spain\label{IAC}}

\author[0000-0002-6424-3410]{Jerome P. de Leon}
\affiliation{Department of Multi-Disciplinary Sciences, Graduate School of Arts and Sciences, The University of Tokyo, 3-8-1 Komaba, Meguro, Tokyo 153-8902, Japan}

\author[0000-0001-6309-4380]{Hiroyuki Tako Ishikawa}
\affiliation{Department of Physics and Astronomy, The University of Western Ontario, 1151 Richmond Street, London, ON N6A 3K7, Canada}

\author[0000-0001-8732-6166]{Joel D. Hartman}
\affil{Department of Astrophysical Sciences, Princeton University, NJ 08544, USA}

\author[0000-0002-4262-5661]{Giuseppe Morello}
\affiliation{Instituto de Astrof\'isica de Andaluc\'ia (IAA-CSIC), Glorieta de la Astronom\'ia s/n, 18008 Granada, Spain}
\affiliation{Instituto de Astrof\'{i}sica de Canarias (IAC), E-38200 La Laguna, Tenerife, Spain\label{IAC}}
\affiliation{Department of Space, Earth and Environment, Chalmers University of Technology, SE-412 96 Gothenburg, Sweden\label{chalmers}}


\author[0009-0002-5067-5463]{Nestor Abreu Garc\'{i}a}
\affiliation{Instituto de Astrof\'{i}sica de Canarias (IAC), E-38200 La Laguna, Tenerife, Spain\label{IAC}}
\affiliation{Departamento de Astrof\'{i}sica, Universidad de La Laguna (ULL), E-38206 La Laguna, Tenerife, Spain\label{ull}}

\author{Leticia \'{A}lvarez Hern\'{a}ndez}
\affiliation{Instituto de Astrof\'{i}sica de Canarias (IAC), E-38200 La Laguna, Tenerife, Spain\label{IAC}}
\affiliation{Departamento de Astrof\'{i}sica, Universidad de La Laguna (ULL), E-38206 La Laguna, Tenerife, Spain\label{ull}}

% ...


\begin{abstract}
Sub-Neptunes with $2-3R_\oplus$ are intermediate in size between rocky planets and Neptune-sized planets.
The orbital properties and bulk compositions of transiting sub-Neptunes provide clues to the formation and evolution of close-in small planets. In this paper, we present the discovery and follow-up of four sub-Neptunes orbiting M dwarfs (TOI-782, TOI-1448, TOI-2120, and TOI-2406), three of which were newly validated by ground-based follow-up observations and statistical analyses.
TOI-782\,b, TOI-1448\,b, TOI-2120\,b, and TOI-2406\,b have radii of $R_\mathrm{p} = 2.740^{+0.082}_{-0.079}\,R_\oplus$, $2.769^{+0.073}_{-0.068}\,R_\oplus$, $2.120\pm0.067\,R_\oplus$, and $2.830^{+0.068}_{-0.066}\,R_\oplus$ and orbital periods of $P = 8.02$, $8.11$, $5.80$, and $3.08$\,days, respectively.
Doppler monitoring with Subaru/InfraRed Doppler instrument led to 2$\sigma$ upper limits on the masses of $<19.1\ M_\oplus$, $<19.5\ M_\oplus$, $<6.8\ M_\oplus$, and $<15.6\ M_\oplus$ for TOI-782\,b, TOI-1448\,b, TOI-2120\,b, and TOI-2406\,b, respectively.
The mass-radius relationship of these four sub-Neptunes testifies to the existence of volatile material in their interiors. 
These four sub-Neptunes, which are located above the so-called ``radius valley'', are likely to retain a significant atmosphere and/or an icy mantle on the core, such as a water world.
We find that at least three of the four sub-Neptunes (TOI-782\,b, TOI-2120\,b, and TOI-2406\,b) orbiting M dwarfs older than 1\,Gyr, are likely to have eccentricities of $e \sim 0.2-0.3$.
The fact that tidal circularization of their orbits is not achieved over 1\,Gyr suggests inefficient tidal dissipation in their interiors.
\end{abstract}


\section{Introduction}\label{sec:intro}

About 70--80\% of stars in the solar neighborhood are M dwarfs \citep[e.g.][]{2019AJ....157..216W,2021AJ....161...63W}.
Nearby M dwarfs are suitable targets to search for low-mass planets.
Recent discoveries of temperate planets orbiting nearby M dwarfs, such as TRAPPIST-1 \citep{2017Natur.542..456G},
provide a unique opportunity for atmospheric characterization of terrestrial planets beyond the solar system.
Near-infrared spectrographs designed for high-precision radial velocity (RV) measurements, such as the Very Large Telescope/CRIRES+ \citep{2016SPIE.9908E..0ID}, the Hobby-Eberly Telescope/Habitable Planet Finder \citep{2018SPIE1mahadevan}, Subaru/InfraRed Doppler \citep{2018SPIE10702E..11K}, CARMENES \citep{2014SPIE.9147E..1FQ}, PARVI \citep{2020JATIS...6a1002G}, NIRPS \citep{2017SPIE10400E..18W},
CFHT/SPIRou \citep{2020MNRAS.498.5684D}, and Gemini/MAROON-X \citep{2018SPIE10702E..6DS}, have begun to find planets down to an Earth mass around M dwarfs, such as Teegarden's Star \citep{2019A&A...627A..49Z}.
Also, space-based planet surveys have accelerated the detection of planets orbiting M dwarfs.
The successor to {\it Kepler}, the Transiting Exoplanets Survey Satellite (TESS), which launched in 2018, has conducted the first-ever all-sky photometric survey for planets around bright stars \citep{2015JATIS...1a4003R}.
The TESS bandpass spanning from 600 to 1000\,nm is well suited for detecting planets around cool stars. About one-fourth of confirmed TESS planets have been discovered around cool stars \citep{2021ApJS..254...39G}.

Both transit detection \citep{2013ApJ...767...95D,2015ApJ...807...45D,2015ApJ...798..112M,2020MNRAS.498.2249H} and RV surveys \citep{2013A&A...549A.109B,2014MNRAS.441.1545T,2020A&A...644A..68M,2021A&A...653A.114S} revealed that small planets are ubiquitous around cool stars. There is a deficit of planets with radii $\sim 1.5-2R_\oplus$, which is called the ``radius valley'', in the planet population observed around FGK-type stars \citep{2018AJ....156..264F,2018MNRAS.479.4786V,2019ApJ...875...29M,2020AJ....160..108B,2022AJ....163..179P}. Also, small planets have a bimodal radius valley distribution in the occurrence rate of planets around M dwarfs \citep{2018AJ....155..127H,2019AJ....158...75H,2020AJ....159..211C}.
The radius valley is interpreted as the compositional transition between bare rocky planets and planets with significant atmospheres \citep{2014ApJ...783L...6W,2015ApJ...801...41R}.
It should be noted that close-in planets ($P < 100$\,days) with radii of $2-4R_\oplus$ above the radius valley (hereafter, sub-Neptunes) outnumber larger planets, such as Neptune- and Jupiter-sized planets \citep[e.g.][]{2021ApJ...911..117S,2022AJ....164..190B}.

The measured densities of almost all of sub-Neptunes 
are too low to be consistent with pure rock and metal, and thereby
suggest a large mass fraction of volatile material.
Many sub-Neptunes with densities lower than those of planets composed of pure H$_2$O retain atmospheres up to the present day. The atmospheres of planets in the proximity of a central star reflect the competing processes of the accumulation of the disk gas \citep[e.g.][]{2012ApJ...753...66I,2014ApJ...797...95L, 2015MNRAS.447.3512O,2018MNRAS.479..635K,2019A&A...623A.179K} and secondary gases, as well as atmospheric escape by photoevaporation \citep[e.g.][]{2013ApJ...776....2L,2013ApJ...775..105O,2017ApJ...847...29O} and core-powered mass loss \citep{2016ApJ...825...29G,2018MNRAS.476..759G,2019MNRAS.487...24G,2022MNRAS.516.4585G}.
Understanding the prevalence of sub-Neptunes with significant atmospheres helps to disentangle the accretion histories of low-mass planets.

We report four sub-Neptunes orbiting M dwarfs from TESS (TOI-782\,b, TOI-1448\,b, TOI-2120\,b, and TOI-2406\,b) validated by ground-based follow-up observations using multicolor photometry, high-resolution imaging, and Subaru/IRD RV measurements and statistical analyses, three of which are newly confirmed by this study; the discovery of TOI-2406\,b was first reported in \citet{2021A&A...653A..97W}.
We also obtain their (upper limit) masses at the $2\sigma$ confidence level from high-precision  RV measurements using Subaru/IRD.
We find that at least three sub-Neptunes around a mature M dwarf with age $\gtrsim 1$\,Gyr are likely to have eccentricities of $e \sim 0.2-0.3$.
Such a nonzero eccentricity of a short-period planet invokes inefficient tidal dissipation in the interior \citep[e.g.][]{2011MNRAS.418.1822W}.

This paper is structured as follows. We describe the TESS photometry and our ground-based follow-up observations in Section \ref{sec:obs}. We present data analyses and the properties of four M dwarfs and their planets in Section \ref{sec:results}. In Section \ref{sec:dis}, we discuss the interior structures of short-period sub-Neptunes with nonzero eccentricities. In the last section, we summarize our results.

\section{Observations}\label{sec:obs}

\subsection{TESS Photometry}

TOI-782 (TIC~429358906), TOI-1448 (TIC~343628284), TOI-2120 (TIC~389900760), and TOI-2406 (TIC~212957629) were observed by TESS in three (Sectors 10, 36, and 63), four (Sectors 15, 16, 17, and 57), five (Sectors 18, 24, 25, 52, and 58), and five (Sectors 3, 30, 42, 43, and 70) sectors, respectively. All of these stars were recorded in postage stamps with a cadence of 2 minutes, except for TOI-1448 in Sector~17 and TOI-2406 in Sector~3 for which only the full-frame images (FFIs) stacked with a cadence of 30 minutes are available.
The data were processed {\bf using} a pipeline developed by the Science Processing Operations Center (SPOC) at NASA Ames Research Center \citep{2016SPIE.9913E..3EJ}, from which transit signals with periods of 8.02, 8.11, 5.80, and 3.08~days were detected in the light curves of TOI-782, TOI-1448, TOI-2120, and TOI-2406, respectively, using dedicated pipelines \citep{2010SPIE.7740E..0DJ,2018PASP..130f4502T}. The transit signatures passed all the diagnostic tests presented in the Data Validation reports, including the odd/even transit depth test, the weak secondary eclipse test, the ghost diagnostic test, the statistical bootstrap test, the period coincidence test with additional transit-like signals, and the difference image centroiding test for the offset from the Tess Input Catalog (TIC) position of the host star. According to the difference image centroiding tests, the host star is located within $6\farcs0 \pm 3\farcs0$, $3\farcs6 \pm 4\farcs2$, $0\farcs7 \pm 3\farcs4$, and $1\farcs6  \pm 3\farcs1$ of the transit signal sources of TOI-782, TOI-1448, TOI-2120, and TOI-2406, respectively. These planetary candidates were released as TESS Objects of Interests (TOIs) 782.01, 1448.01, 2120.01, and 2406.01 by the TESS Science Office  at the Massachusetts Institute of Technology \citep{2021ApJS..254...39G} on UT 2019 June 11, 2019 November 14, 2020 July 19, and 2020 November 24, respectively. \citet{2021A&A...653A..97W} validated the planetary nature of TOI-2406.01, which is now referred to as TOI-2406\,b. In this paper, we validate the planetary nature of the rest, TOI-782.01, TOI-1448.01, and TOI-2120.01, which are hereafter referred to as TOI-782\,b, TOI-1448\,b, and TOI-2120\,b, respectively.

For further analyses, we downloaded the Presearch Data Conditioning Simple Aperture Photometry \citep[PDC-SAP;][]{2012PASP..124.1000S,2012PASP..124..985S,2014PASP..126..100S} where available, otherwise, the TESS-SPOC light curves that are extracted from the FFIs \citep{2020RNAAS...4..201C} in the Mikulski Archive for Space Telescopes at the Space Telescope Science Institute. The light curves were normalized to unity sector by sector after removing data points with bad-quality flags.

\subsection{Multicolor Transit Photometry from the Ground}

We conducted multiband transit photometry of TOI-782\,b, TOI-1448\,b, TOI-2120\,b, and TOI-2406\,b from the ground as part of the TESS Follow-up Observing Program
\citep[TFOP;][]{collins:2019}\footnote{https://tess.mit.edu/followup}
to confirm the transit signals detected by TESS.
We checked for chromaticity in the transit depths, that is, a sign of false positives due to contamination from eclipsing binaries and refined the physical parameters of the planetary candidates.
We utilized a customized version of the {\tt Tapir} software package \citep{Jensen:2013}, the {\tt TESS Transit Finder}, to schedule our transit observations. 
All the observations that are used in our analyses are summarized in Table~\ref{tab:obslog_photometry}. 

Below we briefly describe the observations and data reductions conducted at each facility.

\subsubsection{TCS 1.52m/MuSCAT2}

We observed one transit of TOI-782.01 and one transit of TOI-1448\,b with the multiband imager MuSCAT2 \citep{2019JATIS...5a5001N}, which is mounted on the 1.52m TCS telescope at the Teide Observatory in the Canary Islands, Spain. MuSCAT2 has four optical channels each equipped with a 1k $\times$ 1k CCD camera with a pixel scale of 0\farcs44 pixel$^{-1}$, providing a field of view (FOV) of $7\farcm4 \times 7\farcm4$, and is capable of simultaneous imaging in the $g$, $r$, $i$, and $z_{\rm s}$ bands. The exposure times were set independently for each band as shown in Table~\ref{tab:obslog_photometry}. The airmass range and the median FWHM of the stellar point spread function in each band of each observation are also given in Table~\ref{tab:obslog_photometry}.
We performed image calibration and aperture photometry to extract relative light curves using a pipeline described in \citet{2011PASJ...63..287F}.

\subsubsection{FTN 2m/MuSCAT3}

We observed one transit of TOI-782\,b, one transit of TOI-1448\,b, three transits of TOI-2120\,b, and one transit of TOI-2406\,b with the multiband imager MuSCAT3 \citep{2020SPIE11447E..5KN}, which is mounted on the 2\,m Faulkes Telescope North (FTN), owned and operated by the Las Cumbres Observatory (LCO), located at the Haleakala Observatory in Hawaii, USA. As with MuSCAT2, MuSCAT3 is capable of simultaneous imaging in the $g$, $r$, $i$, and $z_{\rm s}$ bands, but is equipped with a 2k $\times$ 2k CCD camera with a pixel scale of 0\farcs27 pixels$^{-1}$, providing an FOV of $9\farcm1 \times 9\farcm1$, at each channel.
The exposure times, the median FWHM values, and the airmass range for each observation are reported in Table~\ref{tab:obslog_photometry}.
We performed image calibration and aperture photometry on the obtained images in the same way as for the MuSCAT2 data.

\subsubsection{LCO 1m/Sinistro}

We observed six transits of TOI-782\,b, one transit of TOI-1448\,b, and one transit of TOI-2120\,b using the Las Cumbres Observatory Global Telescope \citep[LCOGT;][]{Brown:2013} 1.0\,m network. The telescopes are equipped with $4096\times4096$ Sinistro cameras having an image scale of $0\farcs389$ per pixel, resulting in an FOV of $26\arcmin\times26\arcmin$. The exposure times, the typical FWHM values, and the airmass range for each observation are given in Table~\ref{tab:obslog_photometry}. The images were calibrated using the standard LCOGT {\tt BANZAI} pipeline \citep{McCully:2018} and differential photometric data were extracted using {\tt AstroImageJ} \citep{Collins:2017}.

\subsubsection{TRAPPIST-North}

We observed one full transit of TOI-2120\,b with the 0.6~m TRAPPIST-North telescope located at the Oukaimeden Observatory in Morocco (\citealt{jehin2011,gillon2013,Barkaoui2019}) on UT 2021 July 14. TRAPPIST-North is equipped with a thermoelectrically cooled 2K × 2K Andor iKon-L BEX2-DD CCD camera with a pixel scale of 0.\arcsec64, offering an FOV of 20'× 20'. We collected 186 images in the Sloan $z'$ band with an exposure time of 70\,s. The typical FWHM values and airmass range for this observation are shown in Table~\ref{tab:obslog_photometry}. We performed data reduction and differential aperture photometry using \texttt{prose}\footnote{\url{https://github.com/lgrcia/prose}} \citep{Garcia_2022MNRAS}, which automatically selected the optimum apertures for the photometric data extraction to be 5.05 pixels (3.\arcsec23).

\begin{rotatetable}
\begin{deluxetable*}{lccccccc}
\centerwidetable
\tablenum{1}
\tablecaption{Observation Log of Transit Photometry from the Ground\label{tab:obslog_photometry}}
\tablehead{
\colhead{Planet} & \colhead{Date} & \colhead{Telescope/Site} & \colhead{Instrument} & \colhead{Filter(s)}
& \colhead{Exp. times} & \colhead{FWHM \tablenotemark{a}} & \colhead{Airmass}\\
& (UT) &&&& (s) & ($''$) & (start--min--end)
}
\startdata
TOI-782b & 2020 March 21 & LCO 1m / SSO & Sinistro & $i$ & 60 & 2.2 & 2.00--1.02--1.51\\
TOI-782b & 2021 February 21 & LCO 1m / SSO & Sinistro & $g$ & 270 & 1.6 & 1.54--1.02--1.02\\
TOI-782b & 2021 March 9 & LCO 1m / SSO & Sinistro & $z_{\rm s}$ & 125 & 1.8 & 1.12--1.02--1.19\\
TOI-782b & 2021 May 12 & LCO 1m / SAAO & Sinistro & $g$ & 300 & 2.1 & 1.09--1.03--1.26\\
TOI-782b & 2021 July 15 & LCO 1m / CTIO & Sinistro & $g$ & 300 & 2.3 & 1.10--2.04\\
TOI-782b & 2022 January 8 & FTN 2m / Haleakala & MuSCAT3 & $g$, $r$, $i$, $z_{\rm s}$ & 60, 30, 21, 18 & 2.9, 3.1, 3.1, 3.1 & 2.25--1.30--1.30\\
TOI-782b & 2022 April \bf{30} & LCO 1m / SAAO & Sinistro & $z_{\rm s}$ & 120 & 1.9 & 1.06--1.03--1.41\\
TOI-782b & 2022 April 30 & TCS 1.52m / Teide & MuSCAT2 & $g$, $r$, $i$, $z_{\rm s}$ & 30, 30, 15, 15 & 3.0, 3.0, 2.6, 2.6 & 1.76--1.47--1.51\\
\hline
TOI-1448b & 2020 June 20 & TCS 1.52m / Teide & MuSCAT2 & $g$, $r$, $i$, $z_{\rm s}$ & 120, 60, 20, 15 & 3.0, 2.6, 2.2, 2.1 & 1.55--1.15--1.16\\
TOI-1448b & 2021 July 15 & FTN 2m / Haleakala & MuSCAT3 & $g$, $r$, $i$, $z_{\rm s}$ & 55, 35, 25, 20 & 1.5, 1.3, 1.3, 1.2 & 1.69--1.25--1.41\\
TOI-1448b & 2021 August 24 & LCO 1m / Teide & Sinistro & $i$ & 150 & 1.9 & 1.45--1.15--1.70\\
\hline
TOI-2120b & 2020 January 21 & LCO 1m / McDonald & Sinistro & \bf{$I_{\rm c}$} & 60 & 2.4 & 1.26--1.22--1.33\\
TOI-2120b & 2021 January 16 & FTN 2m / Haleakala & MuSCAT3 & $g$, $r$, $i$, $z_{\rm s}$ & 300, 88, 40, 26 & 2.6, 2.2, 1.7, 1.3 & 1.47--2.13\\
TOI-2120b & 2021 July 14 & TRAPPIST-North / Ouka\"{i}meden & Andor iKon-L & $z$ & 70 & 3.4 & 2.32--1.29\\
TOI-2120b & 2021 August 1 & FTN 2m / Haleakala & MuSCAT3 & $g$, $r$, $i$, $z_{\rm s}$ & 300, 88, 40, 17 & 3.1, 2.8, 2.7, 2.3 & 2.28--1.49\\
TOI-2120b & 2022 January 28 & FTN 2m / Haleakala & MuSCAT3 & $g$, $r$, $i$, $z_{\rm s}$ & 60, 30, 22, 14& 3.0, 3.1, 3.2, 3.0 & 1.48--2.09\\
\hline
TOI-2406b & 2021 August 12 & FTN 2m / Haleakala & MuSCAT3 & $g$, $r$, $i$, $z_{\rm s}$ & 250, 183, 120, 70 & 2.1, 2.1, 2.2, 1.9 & 1.82--1.10--1.13 \\
\enddata
\tablenotetext{a}{Median or typical value during the observation is shown.}
\end{deluxetable*}
\end{rotatetable}

\clearpage

\subsection{High-dispersion Spectroscopy with Subaru/IRD}

We obtained time series of high-resolution spectra of TOI-782, TOI-1448, TOI-2120, and TOI-2406 with the IRD instrument on the 8.2~m Subaru telescope \citep{2012SPIE.8446E..1TT, 2018SPIE10702E..11K} between UT 2019 June 17 and 2022 October 10, as part of the Subaru/IRD TESS intensive follow-up program (ID: S20B-088I). The IRD is a fiber-fed spectrograph covering near-infrared wavelengths from 930 nm to 1740 nm with a spectral resolution of $\approx 70,000$. 
The integration time per exposure was set to 600--1200~s depending on the stellar brightness and observing conditions on each night. In total, we obtained 25, 21, 24, and 22 exposures for TOI-782, TOI-1448, TOI-2120, and TOI-2406, respectively. We also observed at least one telluric standard star (A0 or A1 star) on each night to correct the telluric lines when extracting the template spectrum for the RV analysis. 

The raw IRD data were reduced by the procedure described in \citet{2018SPIE10702E..60K} and \citet{2020PASJ...72...93H}, where the wavelengths were calibrated by spectra from a laser-frequency comb. The reduced one-dimensional spectra have typical signal-to-noise ratios (S/Ns) of 30--60 pixel$^{-1}$ at 1000 nm, except for TOI-2406, whose S/NR was 13--28 due to its faintness. We discarded data with S/Ns lower than 18, which could be affected by detector persistence. We also discarded the data that are logged as being affected by passing clouds or instrumental troubles (e.g., guiding errors) during the exposures, leaving 24, 16, 17, and 11 spectra for TOI-782, TOI-1448, TOI-2120, and TOI-2406, respectively. The RVs for each target were extracted following the procedure of \citet{2020PASJ...72...93H} using the forward modeling technique: for each target, the RV-analysis pipeline derives a telluric-free template spectrum of the target, which is used to model and fit the individual observed spectra taking into account the instantaneous variations in the instrumental profiles. The measured RV values and 1$\sigma$ uncertainties along with the S/N ratios at 1000\,nm of the original spectra are reported in Table~\ref{tab:rv_data}. The median RV internal errors are 5.3, 5.7, 4.6, and 14.5~m~s$^{-1}$ for TOI-782, TOI-1448, TOI-2120, and TOI-2406, respectively. The RV uncertainties vary slightly from frame to frame, reflecting the sky conditions and resulting S/N ratio for each frame. About half of the spectra were obtained under good observing conditions, in which case we achieved the expected S/N ratios and RV internal errors.

\begin{deluxetable}{lcccc}
\tablenum{2}
\tablecaption{Radial Velocities of the Host Stars Measured with Subaru/IRD \label{tab:rv_data}}
\tablewidth{0pt}
\tablehead{
\colhead{Name} & \colhead{Time} & \colhead{RV} & \colhead{$\sigma_{\rm RV}$} & \colhead{S/N \tablenotemark{a}}\\
& (BJD$_{\rm TDB}$) & (m\,s$^{-1}$) & (m\,s$^{-1}$) & 
}
\startdata
TOI-782 & 2458651.807498 & 1.28 & 5.29 & 37.5\\
TOI-782 & 2458651.814354 & -10.92 & 5.77 & 33.3\\
TOI-782 & 2458652.796944 & 0.07 & 5.46 & 37.0\\
TOI-782 & 2458652.804249 & 3.24 & 5.77 & 34.7\\
TOI-782 & 2458652.811564 & 4.3 & 5.55 & 35.9\\
\enddata
\tablecomments{Only a portion of this table is shown here to demonstrate its form and content. A machine-readable version of the full table is available.}
\tablenotetext{a}{Measured at the wavelength of 1000 nm.}
\end{deluxetable}

\subsection{High-resolution Imaging}

To investigate whether there are companion or background stars close to the host stars, we obtained adaptive optics (AO) and speckle images of TOI-782, TOI-1448, and TOI-2120 with various facilities as part of TFOP. The observational dates, instruments, filters, and achieved contrasts are summarized in Table \ref{tab:obslog_imaging}. The observations for individual targets are briefly described below.

\subsubsection{TOI-782}

We conducted AO observations of TOI-782 with the NIRC2 instrument on Keck II behind the natural guide star AO system \citep{2000PASP..112..315W} on UT 2019 June 25 with the $J_{\rm cont}$ ($\lambda_0$ = 1.213 $\mu$m, $\Delta \lambda$ = 0.019  $\mu$m) and Br$\gamma$ ($\lambda_0$ = 2.168 $\mu$m, $\Delta \lambda$ = 0.033 $\mu$m) band filters, and with NIRI \citep{Hodapp2003} on the 8.2\,m Gemini-N telescope on the same night with the Br$\gamma$ ($\lambda_0$=2.1686$\mu$m, $\Delta \lambda$ = 0.0294 $\mu$nm) filter. We also conducted speckle imaging observations with HRCam \citep{2010AJ....139..743T} on the 4.1m SOAR telescope on UT 2019 July 14 through the Cousins $I$-band filter, and used data from the Differential Speckle Survey Instrument (DSSI) \citep{2009AJ....137.5057H} on the Lowell Discovery Telescope (LDT) from UT 2020 February 9 using narrowband filters with center wavelengths of 692nm ($\Delta \lambda$ = 40nm) and 880nm ($\Delta \lambda$ = 50nm), as reported in \citet{2022AJ....163..232C}. We achieved a 5$\sigma$ contrast at a separation of 0\farcs5 of up to 7.7 mag. We did not detect a nearby star within $2\arcsec$ in any of the observations.

\subsubsection{TOI-1448}

We conducted AO observations of TOI-1448 with PHARO \citep{2001PASP..113..105H} behind the natural guide star AO system P3K \citep{dekany2013} of the Palomar 5m telescope on UT 2021 August 24 with a Br$\gamma$-band ($\lambda_0 = 2.1686~\mu$m,  $\Delta\lambda = 0.0326~\mu$m) filter. 
We also conducted speckle imaging observations with NESSI \citep{Scott_2018} on the WIYN 3.5m telescope on UT 2022 September 17 using narrowband filters with central wavelengths of 562nm ($\Delta \lambda$ = 44nm) and 832nm ($\Delta \lambda$ = 40nm). We achieved a 5$\sigma$ contrast at 0\farcs5 of up to 7.7 mag. We did not detect a nearby star within 2\arcsec in any of the observations.

\subsubsection{TOI-2120}

We conducted speckle imaging observations with Speckle Polarimeter \citep{2017AstL...43..344S} on the SAI-2.5m telescope on UT 2020 October 27 in the $I$ band, and with 'Alopeke \footnote{https://www.gemini.edu/sciops/instruments/alopeke-zorro/} \citep{2021FrASS...8..138S} on the Gemini-N telescope on UT 2020 December 3 with the 562nm ($\Delta \lambda$ = 54nm) and 832nm ($\Delta \lambda$ = 40nm) band filters. We achieved a 5$\sigma$ contrast at 0\farcs5 of up to 5.6 mag. We {\bf did not detect a} nearby star within 2\arcsec in any of the observations.

\begin{deluxetable*}{lcccccc}
\tablenum{3}
\tablecaption{Observation Log of High-contrast Imaging\label{tab:obslog_imaging}}
\tablewidth{0pt}
\tablehead{
\colhead{Star} & \colhead{Date} & \colhead{Telescope} & \colhead{Instrument} & \colhead{Filter(s)} & \colhead{Image type} &
\colhead{Contrast at 0\farcs5}\\
& (UT) & & & & & ($\Delta$mag)
}
\startdata
TOI-782 & 2019 June 25 & Keck2 & NIRC2 & $J_{\rm cont}$, Br$\gamma$ & AO & 7.1, 7.7\\
TOI-782 & 2019 June 25 & Gemini-N & NIRI & Br$\gamma$ & AO & 5.6\\
TOI-782 & 2019 July 14 & SOAR & HRCam & $I$ & Speckle & 4.5\\
TOI-782 & 2020 February 9 & LDT & DSSI  & 692nm, 880nm & Speckle & 4.2, 4.6\\
\hline
TOI-1448 & 2021 August 24 & Palomar 5m & PHARO  & Br$\gamma$ & AO & 7.0\\
TOI-1448 & 2022 September 17 & WIYN & NESSI  & 562nm, 832nm  & Speckle & 4.2, 4.0\\
\hline
TOI-2120 & 2020 October 27 & SAI-2.5m & Speckle Polarimeter  & $I$ & Speckle & 3.8\\
TOI-2120 & 2020 December 3 & Gemini-N & 'Alopeke  & 562nm, 832nm & Speckle & 4.3, 5.6\\
TOI-2406 & 2020 December 29 & Gemini-S & Zorro & 562nm, 832nm & Speckle & 4.1, 4.9\\
TOI-2406 & 2021 September 19 & Gemini-S & Zorro & 562nm, 832nm & Speckle & 4.6, 5.3\\
TOI-2406 & 2021 September 19 & Palomar 5m & PHARO & BrG & AO & 5.8\\
\enddata
\tablecomments{SAI: Sternberg Astronomical Institute}
\end{deluxetable*}

\section{Data Analysis and Results}\label{sec:results}

\subsection{Properties of the Host Stars}

\subsubsection{Spectroscopic Properties}
\label{sec:spectroscopic_properties}

We estimated the metallicity ([Fe/H]) and the effective temperature ($T_{\mathrm{eff}}$) of four host stars using the IRD spectra following the procedure described in \citet{2022Mori}. For TOI-782b, TOI-1448b, and TOI-2120b, we used the template spectra identical to those used for the RV analysis, but for TOI-2406b, we used a combined spectrum before deconvolving the instrument profile to avoid amplifying the additional noise due to the low S/N. The basic concept of our analysis is an equivalent width (EW) comparison of individual atomic and molecular absorption lines between synthetic and observed spectra. To determine the $T_{\mathrm{eff}}$ value, 47 FeH molecular lines at $990-1012$\,nm were used (see \citet{2022AJ....163...72I} for details). We employed a total of 25 atomic lines corresponding to Na I, Mg I, Ca I, Ti I, Cr I, Mn I, Fe I, and Sr II as a measure of the elemental abundances \citep{2020PASJ...72..102I}. We selected only atomic lines with a high S/N. We obtained the EWs of the atomic lines by fitting the synthetic spectra line by line, while the EWs of the FeH lines were obtained from a Gaussian fitting to each line. The logarithm of the surface gravity $\log g$ and the microturbulent velocity were fixed at 5.0 and 0.5 km s$^{-1}$, respectively, for all the targets. We checked that this assumption causes a negligible change in the final $T_{\mathrm{eff}}$ and [Fe/H] estimates.

First, we derived a $T_{\mathrm{eff}}$ from the EWs of the FeH lines, assuming solar abundances. Subsequently, we determined the individual abundances of the eight elements, i.e., [X/H], using the $T_{\mathrm{eff}}$ value. Then, we carried out these analyses for [X/H] and $T_{\mathrm{eff}}$ in an iterative process until the $T_{\mathrm{eff}}$ and [Fe/H] matched the values assumed in the previous iteration step within their uncertainties. We obtained $[\mathrm{Fe}/\mathrm{H}] =0.34 \pm 0.15$, $0.37 \pm 0.14$, $0.39 \pm 0.22$, and $-0.26 \pm 0.24$ dex, and $T_{\mathrm{eff}} =$ $3379 \pm 100$\,K, $3390 \pm 100$\,K, $3183 \pm 100$\,K, and $3308 \pm 100$\,K for TOI-782, TOI-1448, TOI-2120, and TOI-2406, respectively. Note that our [Fe/H] value for TOI-2406 is consistent with that derived by \citet{2021A&A...653A..97W} within the uncertainties. 

Using the IRD spectra, we also derived the systemic RVs of our targets. We fitted Gaussian functions to relatively deep atomic lines ($30-40$ lines) at near-infrared wavelengths and compared the line centers with their vacuum wavelengths. The systemic RVs and their uncertainties were then derived from the mean values and the scatter of the measurements as listed in Table \ref{tab:star}. Inputting the systemic RVs together with the coordinates, parallaxes, and proper motions, we computed the Galactic space velocities $(U,\,V,\,W)$ for the four stars using the Python script {\tt uvw\_errs} \citep{2016zndo....192159R}. The derived $UVW$ velocities are also listed in Table \ref{tab:star}. For all of our targets, we found that the sky-projected rotational velocity ($v \sin i_s$) of each star is too small (smaller than the spectral resolution, or $\sim$4.3 km\,s$^{-1}$) to detect from IRD spectra in a well-calibrated way.

\subsubsection{Photometric Properties}
\label{sec:photometric_properties}

We estimated the stellar parameters other than [Fe/H] mainly from the photometric magnitudes as follows. First, we estimated the stellar radius ($R_s$) and mass ($M_s$) through the empirical luminosity-metallicity-radius and luminosity-metallicity-mass relations of \citet{2015ApJ...804...64M} and \citet{2019ApJ...871...63M}, respectively, adopting the parallaxes from Gaia~DR3 \citep{2016A&A...595A...1G,2023A&A...674A...1G}; assuming the distance $d$ is the inverse of the parallax, the $K_s$-band magnitudes from the Two Micron All Sky Survey (2MASS;\citep{2006AJ....131.1163S}), and the [Fe/H] values derived in Section~\ref{sec:spectroscopic_properties}. The stellar radii were estimated to be $0.398 \pm 0.012$\,$R_\odot$, $0.376 \pm 0.011$\,$R_\odot$, $0.237 \pm 0.007$\,$R_\odot$, and $0.201 \pm 0.006$\,$R_\odot$ for TOI-782, TOI-1448, TOI-2120, and TOI-2406, respectively. These values were used as priors in the spectral energy distribution (SED) fitting as described below. The stellar masses were estimated to be $0.397 \pm 0.010$\,$M_\odot$, $0.372 \pm 0.009$\,$M_\odot$, $0.211 \pm 0.005$\,$M_\odot$, and $0.166 \pm 0.004$\,$M_\odot$ for TOI-782, TOI-1448, TOI-2120, and TOI-2406, respectively. We adopted these values as our fiducial values for $M_s$, which are listed in Table~\ref{tab:star}.

Next, we updated $T_\mathrm{eff}$, $R_s$, and $d$ by fitting a stellar-atmospheric model to the SED constructed from the photometric magnitudes. For the photometric magnitudes, we used the catalog magnitudes in the $BP$, $G$, $RP$ bands from Gaia~DR3, $J$, $H$, $K_s$ bands from 2MASS, and W1, W2 from AllWISE \citep{2010AJ....140.1868W,2011ApJ...731...53M} for all host stars except for TOI-2120. For TOI-2120, because there is a nearby background star (TIC\,389900766) that was separated from TOI-2120 by $\sim$9$^{''}$, $\sim$6$^{''}$, and $\sim$4$^{''}$ at the times when the data were collected by 2MASS, AllWISE, and Gaia, respectively, we did not use the W1 and W2 magnitudes from AllWISE which could be contaminated by the flux from this nearby star.
For the fitting, the tabulated BT-Settl model \citep{2011ASPC..448...91A} was linearly interpolated to calculate a synthetic SED for given $T_\mathrm{eff}$, [Fe/H], and $\log g$. We performed a Markov Chain Monte Carlo (MCMC) analysis using {\tt emcee} \citep{2013PASP..125..306F} to calculate the posterior probability distributions of $T_\mathrm{eff}$, $R_s$, and $d$, where we applied normal priors to $T_\mathrm{eff}$, $R_s$, and $d$ with the mean and standard deviation estimated from IRD, the empirical relation, and Gaia~DR3, respectively. The [Fe/H] and $M_s$ values are taken from Gaussian distributions with the mean and standard deviation estimated from IRD and the empirical relation. We calculated $\log g$ from the given $R_s$ and $M_s$. A single jitter term was added to all the magnitude errors in quadrature and left free in the MCMC analysis. Note that we assumed no interstellar extinction, which is negligible for all the four stars according to the 3D extinction map of \cite{2019ApJ...887...93G}.\footnote{\url{http://argonaut.skymaps.info}}
As a result, we obtained the values of $T_\mathrm{eff}$ to be $3370\ ^{+29}_{-32}$\,K, $3412\ ^{+28}_{-32}$\,K, $3131 \pm 30$\,K, and $3100\ ^{+31}_{-24}$\,K, and values of $R_s$ of $0.413 \pm 0.008$\,$R_\odot$, $0.380 \pm 0.007$\,$R_\odot$, $0.245 \pm 0.006$\,$R_\odot$, and $0.204 \pm 0.004$\,$R_\odot$ for TOI-782, TOI-1448, TOI-2120, and TOI-2406, respectively. The updated values of $T_\mathrm{eff}$ are consistent with those derived from the IRD spectra (Section~\ref{sec:spectroscopic_properties}) within 1$\sigma$ for TOI-782, TOI-1448, and TOI-2120 and within 2$\sigma$ for TOI-2406. The updated values of $R_s$ are all within 1$\sigma$ of those derived from the empirical relation.
Our values of $T_\mathrm{eff}$, $R_s$, and $d$ are listed in Table~\ref{tab:star}. The prior and posterior SED models along with the fitted SED data are shown in Figure \ref{fig:SED}.

\begin{deluxetable*}{lccccccr}
\tablenum{4}
\tablecaption{Properties of the Four M Dwarf Hosts\label{tab:star}}
\tablewidth{0pt}
\tablehead{
\colhead{Parameters} &  \colhead{Unit}  & \multicolumn5c{Values}\\
\colhead{} & \colhead{} & \colhead{TOI-782} & \colhead{TOI-1448} & \colhead{TOI-2120} & \multicolumn2c{TOI-2406}\\
\colhead{} &  \colhead{} &  \colhead{}  & \colhead{}  & \colhead{} &  \colhead{This work}  &   \colhead{Wells et al.(2021)}
}
\startdata
TIC & & 429358906 & 343628284 & 389900760 & \multicolumn2c{212957629}\\
Gaia DR3 ID & & 3518374197418907648 & 2189711770761074816 & 513299860904522752 & \multicolumn2c{2528453161326406016} & \\
RA \tablenotemark{a} & & 12:15:40.8810 & 21:5:17.7597 & 01:34:1.7802 & \multicolumn2c{00:35:13.4606}\\
Dec \tablenotemark{a} & & -18:54:37.234 & +57:46:15.908 & +65:30:50.248 & \multicolumn2c{-03:22:19.668}\\
pmRA \tablenotemark{a} & mas yr$^{-1}$ & $-163.739 \pm 0.023$ & $-15.617 \pm 0.020$ & $306.045 \pm 0.014$ & \multicolumn2c{$226.011 \pm 0.052$}\\
pmDec \tablenotemark{a} & mas yr$^{-1}$ & $-36.597 \pm 0.020$ & $-64.690 \pm 0.019$ & $-215.453 \pm 0.019$ & \multicolumn2c{$-336.242 \pm 0.035$}\\
Parallax \tablenotemark{a} & mas & $19.153 \pm 0.020$ & $13.585 \pm 0.015$ &  $31.072 \pm 0.018$& \multicolumn2c{$17.985 \pm 0.041$}\\
\\
$T$ (TESS) & mag & $12.2875 \pm 0.0073$ & $13.1717 \pm 0.0074$ & $12.8233 \pm 0.0074$ & \multicolumn2c{$14.3109 \pm 0.0074$}\\
$G$ (Gaia) & mag & $13.5428 \pm 0.0004$ & $14.3794 \pm 0.0006$ & $14.1593 \pm 0.0005$ & \multicolumn2c{$15.6630 \pm 0.0006$}\\
$BP$ (Gaia) & mag & $15.0559 \pm 0.0016$ & $15.7910 \pm 0.0030$ & $15.9017 \pm 0.0028$ & \multicolumn2c{$17.4339 \pm 0.0060$}\\
$RP$ (Gaia) & mag & $12.3426 \pm 0.0007$ & $13.2132 \pm 0.0013$ & $12.9023 \pm 0.0009$ & \multicolumn2c{$14.3930 \pm 0.0012$}\\ 
$J$ (2MASS) & mag & $10.675 \pm 0.021$ & $11.599 \pm 0.030$ & $11.100 \pm 0.023$ & \multicolumn2c{$12.633 \pm 0.024$}\\
$H$ (2MASS) & mag & $10.108 \pm 0.024$ & $11.033 \pm 0.030$ & $10.452 \pm 0.026$ & \multicolumn2c{$12.129 \pm 0.024$}\\
$K$ (2MASS) & mag & $9.860 \pm 0.024$ & $10.765 \pm 0.020$ & $10.205 \pm 0.023$ & \multicolumn2c{$11.894 \pm 0.025$}\\
$W1$ (WISE) & mag & $9.717 \pm 0.022$ & $10.628 \pm 0.022$ & --- & \multicolumn2c{$11.706 \pm 0.023$}\\
$W2$ (WISE) & mag & $9.559 \pm 0.020$ & $10.509 \pm 0.020$ & --- & \multicolumn2c{$11.458 \pm 0.022$}\\
\\
Mass & $M_\odot$ & $0.397\pm0.010$ & $0.372\pm0.009$ & $0.211\pm0.005$ & $0.166\pm0.004$ & $0.162\pm0.0008$  \\
Radius & $R_\odot$  & $0.413\pm0.008$ & $0.380\pm0.007$ & $0.245\pm0.006$ & $0.204\pm0.004$  &  $0.202\pm0.011$\\
Effective temp. & K  & $3370\ ^{+29}_{-32}$ & $3412\ ^{+28}_{-32}$ & $3131\pm30$ & $3100\ ^{+31}_{-24}$ &  $3100\pm75$\\
$\log{g}$ & cm\,s$^{-2}$ & $4.804\pm0.021$ & $4.849\pm0.019$
& $4.985\pm0.024$ & $5.041\ ^{+0.022}_{-0.020}$  & $5.037^{+0.053}_{-0.051}$ \\
Density & g\,cm$^{-3}$ & $7.9\ ^{+0.6}_{-0.5}$ & $9.5\ ^{+0.6}_{-0.5}$ & $20.3\ ^{+1.7}_{-1.5}$ & $27.7\ ^{+2.1}_{-1.7}$  & $27.7^{+5.3}_{-4.2}$ \\
Luminosity & $L_\odot$  & $0.01984\ ^{+0.00044}_{-0.00052}$ & $0.01758\ ^{+0.00030}_{-0.00033}$ & $0.00517\ ^{+0.00013}_{-0.00016}$ & $0.00345\ ^{+0.00008}_{-0.00009}$ &  $0.0034\pm0.00016$  \\
Distance & pc & $52.210\pm0.055$ & $73.610\pm 0.079$ & $32.183\pm 0.018$ & $55.60\pm0.12$ &  $55.6\pm0.13$  \\
Rot. period & d & $65.257 \pm 0.063$ & $42.28 \pm 0.085$ & -- & -- & --\\
$[\mathrm{Fe}/\mathrm{H}]$ & dex & $0.34\pm0.15$ & $0.37\pm0.14$ & $0.39\pm0.22$ & $-0.26\pm0.24$  & $-0.38\pm0.07$ \\
$[\mathrm{Na}/\mathrm{H}]$ & dex & $0.24\pm0.14$ & $0.31\pm0.17$ & $0.31\pm0.25$ & $-0.21\pm0.28$  & -- \\
$[\mathrm{Mg}/\mathrm{H}]$ & dex & $0.45\pm0.19$ & $0.26\pm0.18$ & $0.22\pm0.30$ & $-0.16\pm0.39$  & -- \\
$[\mathrm{Ca}/\mathrm{H}]$ & dex & $0.39\pm0.15$ & $0.18\pm0.16$ & $0.12\pm0.27$ & $-0.44\pm0.35$  & -- \\
$[\mathrm{Ti}/\mathrm{H}]$ & dex & $0.80\pm0.30$ & $0.66\pm0.26$ & $0.71\pm0.38$ & $-0.06\pm0.31$  & -- \\
$[\mathrm{Cr}/\mathrm{H}]$ & dex & $0.40\pm0.16$ & $0.35\pm0.15$ & $0.32\pm0.22$ & $-0.38\pm0.18$  & -- \\
$[\mathrm{Mn}/\mathrm{H}]$ & dex & $0.36\pm0.23$ & $0.38\pm0.22$ & $0.17\pm0.36$ & $-0.61\pm0.41$  & -- \\
$[\mathrm{Sr}/\mathrm{H}]$ & dex & $0.54\pm0.22$ & $0.55\pm0.20$ & $0.40\pm0.30$ & $-0.54\pm0.37$  & -- \\
\\
Systemic RV & km\,s$^{-1}$ & $-7.8 \pm 0.1$ & $ -23.7 \pm 0.1$ & $20.9 \pm 0.1$ & $-29.5 \pm 0.1$ & --\\  
$U$ & km\,s$^{-1}$ & $-33.57\pm0.13$ & $+22.78\pm0.05$ & $-54.39\pm0.09$ & $-2.35\pm0.06$& $+1.5\pm0.9$\\
$V$ & km\,s$^{-1}$ & $-19.27\pm0.12$  & $-19.79\pm0.10$ & $-13.60\pm0.09$ & $-109.69\pm0.44$ & $-93.2\pm2.1$  \\
$W$ & km\,s$^{-1}$ & $-17.54\pm0.09$  & $-13.71\pm0.04$  & $-23.60\pm0.04$ & $-12.39\pm0.20$  & $-46.4\pm5.2$\\
\enddata
\tablenotetext{a}{From Gaia DR3. The values are given for epoch=J2016.0.}
\end{deluxetable*}

\subsubsection{Stellar Activity and Rotation}

\begin{figure}[ht]
    \gridline{\fig{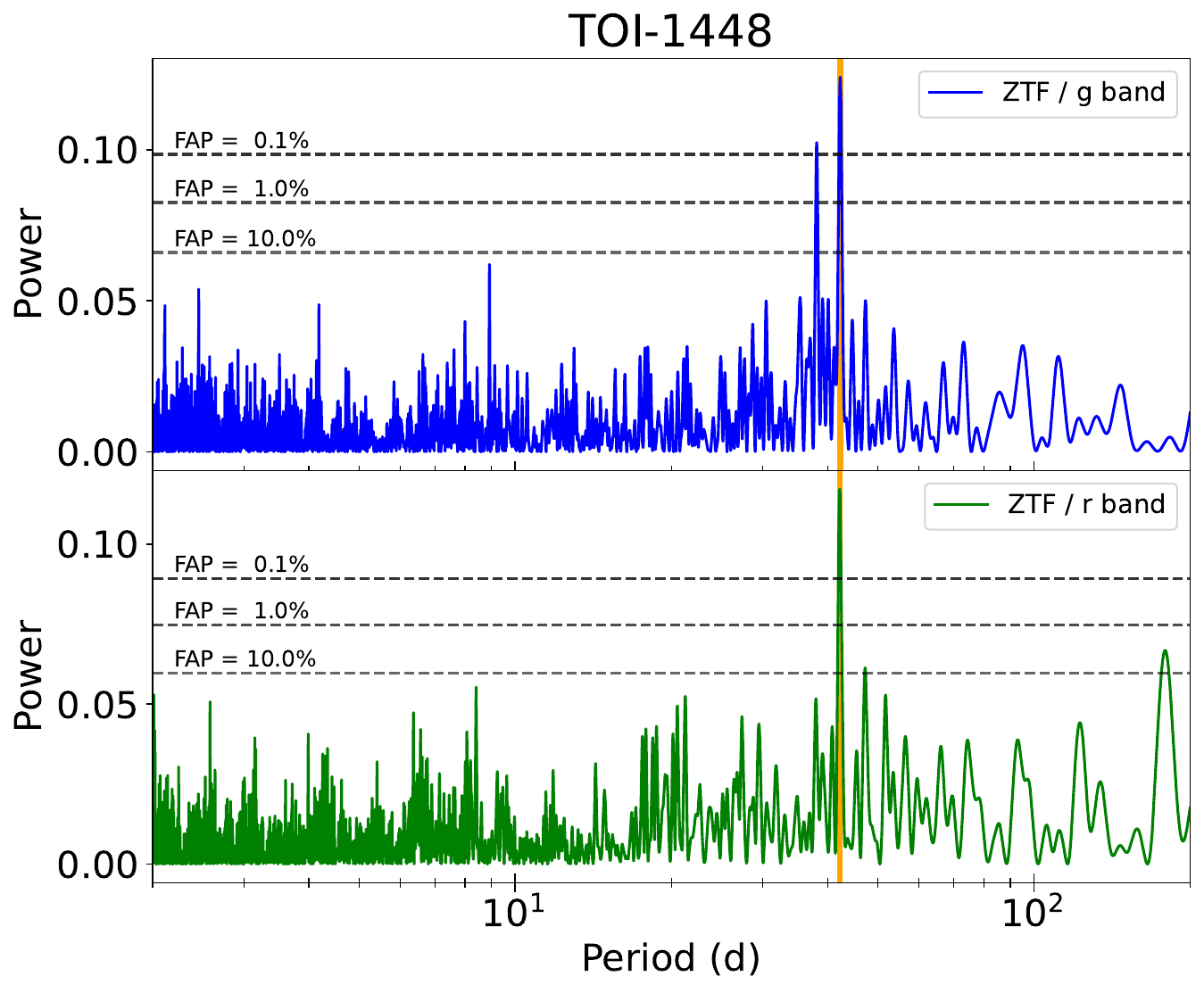}{0.5\textwidth}{(a)}}
    \gridline{\fig{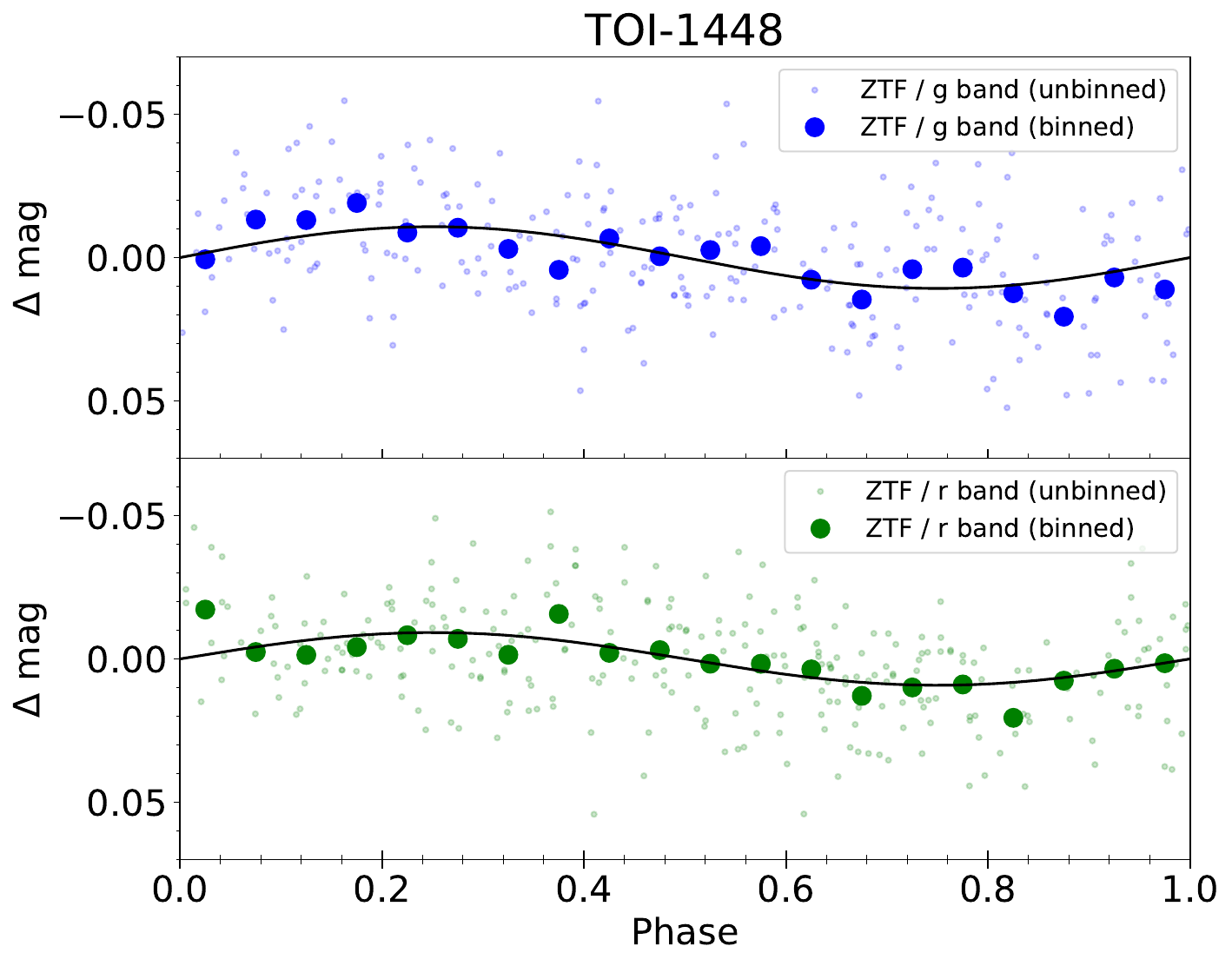}{0.5\textwidth}{(b)}}
	\caption{(a) Power spectra for the $g$- (top) and $r$-band  (bottom) ZTF light curves of TOI-1448. The gray dashed lines indicate FAPs of 0.1, 1, and 10\% from top to bottom. Significant, coincident signals as shown by the orange vertical lines) can be seen at a period of 42.3 days. (b) The ZTF light curves of TOI-1448 in the $g$ band (top) and the $r$ band (bottom) folded with the peak periods. Light dots and dark circles are unbinned and binned data, respectively. Black curves show the best-fit sine function.
	\label{fig:ZTF_TOI-1448}
	}
\end{figure}

In order to investigate the magnetic activity and rotation periods of the host stars, we searched for long-term photometric variability of the host stars in the TESS light curves. We found no significant variability in any of the four stars. We also searched for periodic variability in the $g$- and $r$-band archival light curves of the Zwicky Transient Facility (ZTF) DR16\footnote{https://irsa.ipac.caltech.edu/Missions/ztf.html} which are available for TOI-1448, TOI-2120, and TOI-2406, and in the $g$- and $V$-band light curves from ASAS-SN\footnote{https://asas-sn.osu.edu/} for TOI-782, by performing a Lomb-Scargle periodogram analysis. As a result, there were no plausible periodic signals in any of the light curves except for the $g$- and $r$-band ZTF light curves of TOI-1448, in which we found significant and coincident periodic signals with false alarm probabilities (FAPs) of less than 0.1\%, as shown in Figure~\ref{fig:ZTF_TOI-1448}. The strongest signals in the power spectra for the $g$- and $r$-bands are present at periods of $42.32 \pm 0.11$~d and $42.22 \pm 0.13$~days, having variability amplitudes of $10.8 \pm 1.8$~mmag and $9.2 \pm 1.5$~mmag, respectively. Considering the significance and period coincidence of the two signals, we attribute these signals to stellar rotation. The weighted mean of the two periods is calculated to be $42.28 \pm 0.085$~days, which is listed in Table~\ref{tab:star}.

\begin{figure*}[ht!]
	\epsscale{1.2}
	\plotone{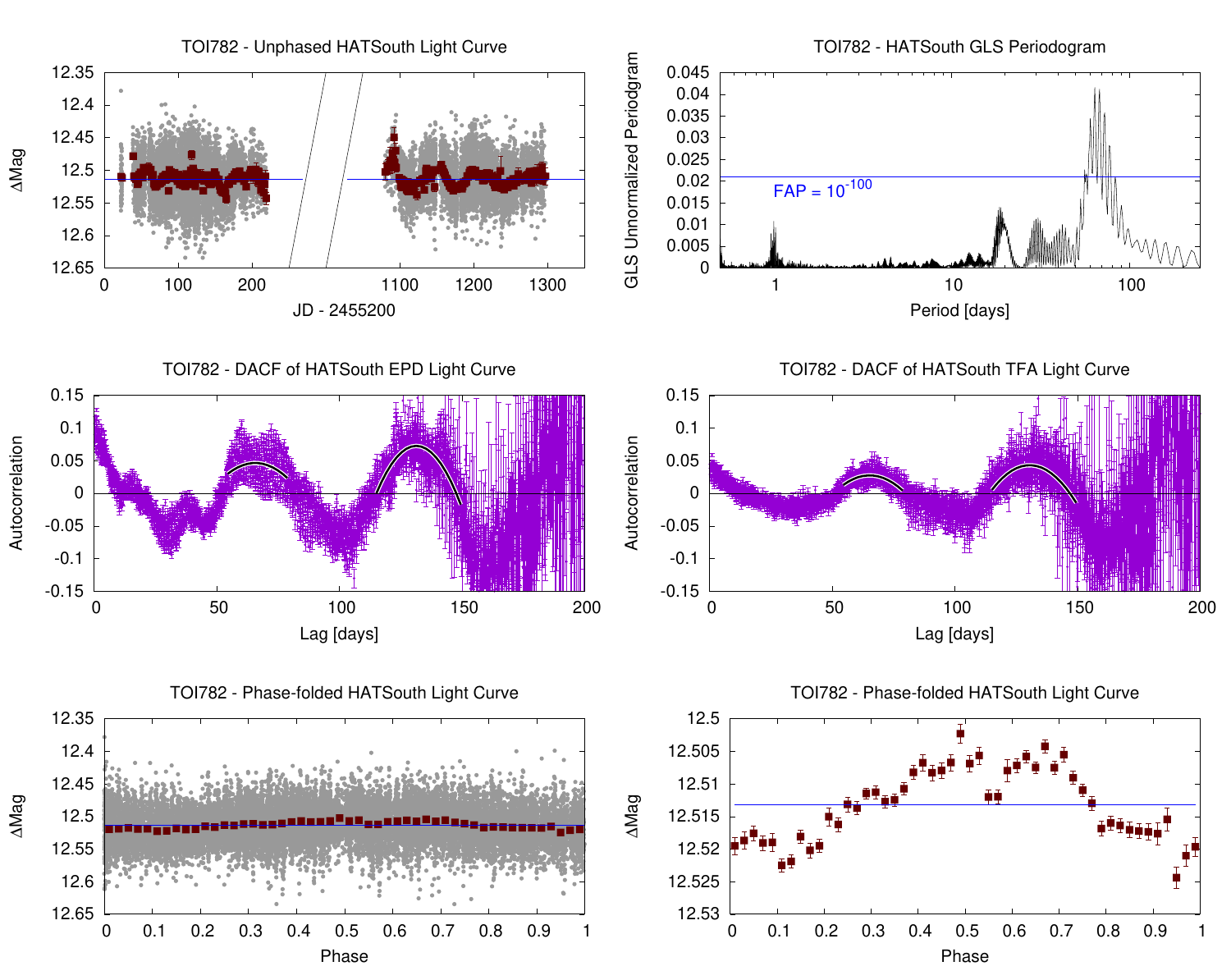}
	\caption{{\em Top left:} the HATSouth light curve of TOI-782 plotted against time. The grey points show the unbinned data, while the red squares show the data binned by 1\,day. For clarity, we cut out a gap in the data as shown. The light curve shown in this figure has had an EPD detrending applied to it, but not the more aggressive TFA detrending that is typically also applied to HATSouth light curves. {\em Top right:} The GLS unnormalized periodogram of the HATSouth light curve of TOI-782 shows a strong peak at a period of $P = 64$\,days. {\em Middle left:} The DACF of the HATSouth EPD light curve. The first and second prominent peaks are fitted with parabolas to determine the period.
    {\em Middle right:} same as on the left, but here we show the DACF of the HATsouth TFA light curve. Some shorter-lag variations in the EPD DACF are not present, suggesting that they are likely due to instrumental systematics.
    {\em Bottom left:} the HATSouth EPD light curve phase folded at the detected period. The gray-scale points show the unbinned data, while the red squares show the phase-binned data with a bin size of 0.02. {\em Bottom tight:} same as on the left, but here we show only the phase-binned data and zoom in to see the detected photometric variation more clearly.
	\label{fig:HS_TOI-782}
	}
\end{figure*}

TOI-782 was observed by the HATSouth ground-based telescope network \citep{2013PASP..125..154B} during two observing seasons between 2010 January 26 and 2013 July 23, with a gap in the observations between 2010 August 10 and 2012 December 18. A total of 25,247 photometric observations were obtained in the $r$-band with an exposure time of 4 minutes. The observations were reduced to trend-filtered light curves following the procedures described in \citet{2013PASP..125..154B}. We used the generalized lomb-scargle periodogram (GLS) \citep[][]{2009A&A...496..577Z} to detect a significant periodic signal in the light curve at a period of $P = 64$\,days, and with a semiamplitude of $7.28 \pm 0.23$\,mmag, assuming a strictly periodic sinusoidal signal. The signal has a formal bootstrap-calibrated formal FAP of less than $10^{-200}$, with the number of independent trials calibrated through a bootstrap procedure. This indicates a vanishingly small chance of white noise data producing a periodogram peak of this height. However, the signal also appears to vary over time. When we analyze the two seasons separately, the peak period in the GLS periodogram for the first season is at a period $\sim 30$\,days, with a lower significance, while the second season shows a strong peak at $\sim 60$\,days. The semi-amplitude of the $P = 64$\,days signal is also lower in the first season, with a value $\sim 2$\,mmag. Because of the apparent evolution of the signal with time, we also calculated the Discrete Autocorrelation Function \citep[DACF;][]{1988ApJ...333..646E} of the light curve, and find a clear cyclic variation with a period of $P = 65.653 \pm 0.051$\,days based on fitting two parabolas to the first and second prominent peaks in the DACF. Some less prominent peaks also appear at shorter time lags. The analysis discussed so far is for a partial detrending of the light curve against a set of auxiliary instrumental parameters (henceforth referred to as external parameter decorrelation (EPD)), which yields a point-to-point RMS of 25\,mmag. When we apply the more aggressive trend-filtering algorithm \citep[TFA;][]{2005MNRAS.356..557K} to the HATSouth light curve, which reduces the RMS to 20\,mmag, the signal remains at a high level of significance, but the amplitude of this long-period variation is reduced to $4.18 \pm 0.22$\,mmag. The DACF of the TFA light curve appears to show a cleaner cyclic variation, with the lag between the first and second peaks indicating a period of $P = 65.257 \pm 0.063$\,days. This suggests that the additional less pronounced, shorter-lag peaks present in the EPD DACF are likely due to instrumental artifacts. The periodogram, DACF and light curve are shown in Figure~\ref{fig:HS_TOI-782}. 

\subsection{Stellar Age}
TOI-782 and TOI-1448 have a slow rotation with $P \sim 65.7$\,days and $42.3$\,days, respectively. Neither TOI-782 nor TOI-2120 show any activity variations due to stellar spots.
The low-velocity stars, TOI-782, TOI-1448, and TOI-2120, are probably thin disk stars (see Table\,\ref{tab:star}), whereas the total velocity of the metal-poor TOI-2406, $v_\mathrm{tot} = \sqrt{U^2 + V^2 + W^2} \sim110$\,km\,s$^{-1}$, suggests that it is a member of the thick disk \citep[see also][]{2021A&A...653A..97W}.
In fact, the thick disk to thin disk probability ratios of TOI-782, TOI-1448, TOI-2120, and TOI-2406 are estimated to be $0.0125, 0.0144, 0.0213,\,\mathrm{and}\,38.9$, respectively, from their Galactic kinematics \citep{2014A&A...562A..71B}, respectively. Also, the BANYAN $\Sigma$ analysis \citep{2018ApJ...856...23G} confirm that four these M dwarfs are not bona fide members of the young association within $\sim$100\,pc.

\subsection{Validation of the Planetary Candidates}
\label{sec:validation}

In the following paragraphs, we qualitatively eliminate the classes of false positive scenarios that can mimic the transit signal including an EB with a grazing transit geometry, a hierarchical eclipsing binary (HEB), and a diluted eclipse of a background (or foreground) eclipsing binary (BEB) along the line of sight of the target. Although the planetary nature of TOI-2406\,b has already been validated by \cite{2021A&A...653A..97W}, we will still validate all four planetary candidates for completeness.

First, the renormalised unit weight error (RUWE) from Gaia DR3\footnote{https://gaia.ari.uni-heidelberg.de/singlesource.html} is 1.25, 1.08, 1.15, and 1.06 for TOI-782, TOI-1448, TOI-2120, and TOI-2406, respectively, which are all low enough to be consistent with a single star being the source of the astrometric motion \citep{2020Belokurov}. We also rule out the EB scenario based on the mass constraint derived in Section~\ref{sec:joint}. Moreover, we do not detect any wavelength dependence of the transit depth from our chromatic transit analysis (Section \ref{sec:trfit_ground}), which is consistent with the absence of contamination from a star of a spectral type (color) different from the host star.
In the absence of dilution, the measured planetary radii become significantly smaller than the lower limit of 0.8$R_{\rm{Jup}}$ expected for brown dwarfs \citep{2011Burrows}. Grazing transit geometries can be eliminated, as the impact parameter is constrained to $b<0.6$ at the 99\% level based on our transit and contamination analyses. The apparent boxy shapes of the transit light curves are in stark contrast with the V-shaped transit expected for grazing orbits. Hence, the grazing EB scenario is ruled out.

Second, we constrain the classes of HEBs that reproduce the observed transit depth and shape using our multiband observations. We aim to compute the eclipse depths for a range of plausible HEBs in the bluest and reddest bandpasses where they are expected to vary significantly.
We adopt the method presented in \citet{2022Mori}, which is based on \citet{2020BoumaTOI837}, to perform the calculation over a grid of eclipsing companions.\footnote{We assumed cases where the impact parameter is allowed to vary from zero to one.} Comparing the simulated eclipse depth with the observed one (Section \ref{sec:trfit_ground}) in each band, we find that no plausible HEB configuration explored in our simulation can reproduce the observed transit depths in the multiple bands simultaneously as well as the transit shape. 
To check the possibility of stellar companions spectroscopically, we visually searched for secondary peaks in the stellar line profiles. For all the four targets, the mean line profile, a cross-correlation function of the IRD spectra against a template spectrum, exhibited no sign of a stellar companion with a contrast higher than $\sim 0.1$ in the near-infrared wavelengths. 
Hence, the HEB scenario is ruled out.

Third, the probability of being a BEB is low a priori because our targets are far from the Galactic plane, except for TOI-1448 and TOI-2120,\footnote{TOI-1448 and 2120 have Galactic latitudes of $b$=+7.$^\circ$1234 and $b$=+3.$^\circ$007295, respectively.} and 
we argue below that the BEB scenario for TOI-1448 and 2120 is very unlikely. 
First, there is a background star (TIC\,389900766; $\Delta G$=5.1) near TOI-2120 with a separation of 2\farcs2--2\farcs5 at the times of the ground-based transit observations, which contaminates the photometric apertures for these observations (3$''$ or larger in radius). However, we did not detect any significant chromatic difference in the transit depth in the analysis of the ground-based transit light curves (Section \ref{sec:trfit_ground}), which cannot be explained by the BEB scenario given the chromatic flux ratios of the background star to TOI-2120 of 1.3\%, 1.6\%, 0.78\%, and 0.79\% for the $g$, $r$, $i$, and $z_s$ bands, respectively, measured from the Pan-STARRS1 catalog \citep{2016arXiv161205560C}.
Next, our high-resolution speckle and AO imaging ruled out any other nearby star and blended sources down to 0\farcs5 at a $\Delta$mag of 7.7 (2166\,nm), 7.0 (2166\,nm), 5.6 (832\,nm), and 5.2 (832\,nm) for TOI-782, TOI-1448, TOI-2120 and TOI-2406, respectively. 
We statistically estimate the probability of a chance-aligned star, using the population synthesis code {\tt TRILEGAL}\footnote{http://stev.oapd.inaf.it/cgi-bin/trilegal} \citep{2005GirardiTrilegal}, which simulates the Galactic stellar population along any line of sight. Given the positions of our targets, we found probabilities of $P_{\rm{BEB}}<10^{-6}$ to find a star brighter than $T_{\rm{mag}}\simeq$16,\footnote{$T_{\rm{mag}}$ denotes the TESS magnitude. The maximum $\Delta$magnitude was computed using $dT=-2.5 \log_{10}$(depth), which translates to the magnitude that can produce a 100\% eclipse.} within 0\farcs5.
Assuming that all such stars are binaries and preferentially oriented edge on to produce eclipses with the period and depth consistent with the TESS detection, this can represent a conservative upper limit to the BEB scenario.

Finally, we quantify the false positive probability (FPP) of our targets using the Python package {\tt TRICERATOPS} \citep{2020GiacaloneDressing}. We used the ephemerides and depths reported in Table~\ref{tab:joint}, the contrast curves from the high-resolution images with the highest contrast achieved, and the MuSCAT3 $i$-band  follow-up light curves as inputs to {\tt TRICERATOPS}. 
Although we were able to rule out the classes of EB, BEB, and HEB, we ran {\tt TRICERATOPS} considering all of these scenarios for completeness. 
We obtained formal FPPs of the order of $10^{-5}$, $10^{-4}$, $10^{-7}$, and $10^{-5}$
for TOI-782\,b, TOI-1448\,b, TOI-2120\,b, and TOI-2406\,b, respectively. 
These values are well below the threshold of FPP$< 1.5$\% 
prescribed by \citet{2021GiacaloneTOI} which statistically validates them as bona fide planets.

\begin{figure*}[ht]
    \gridline{\fig{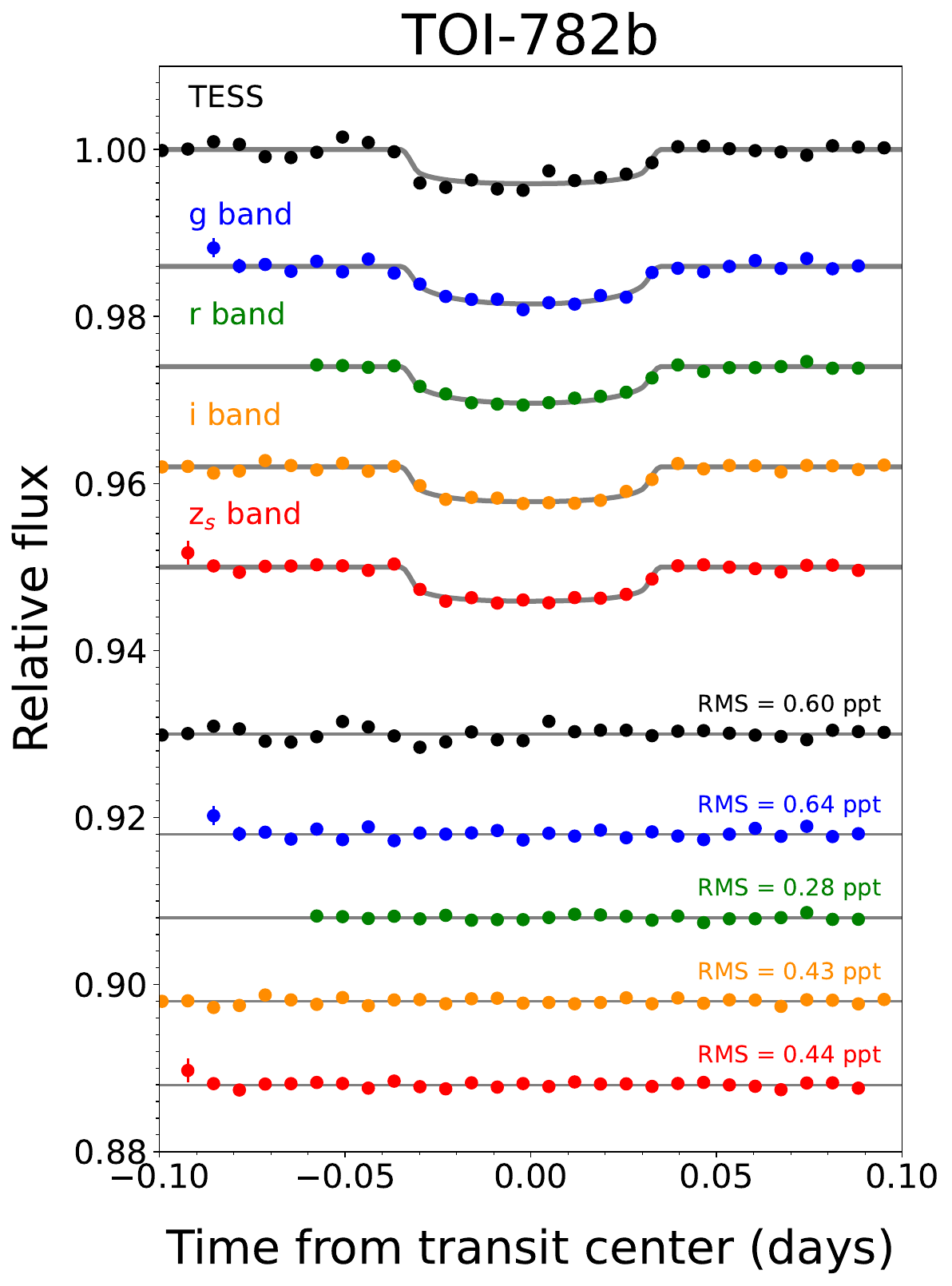}{0.5\textwidth}{(a)}
              \fig{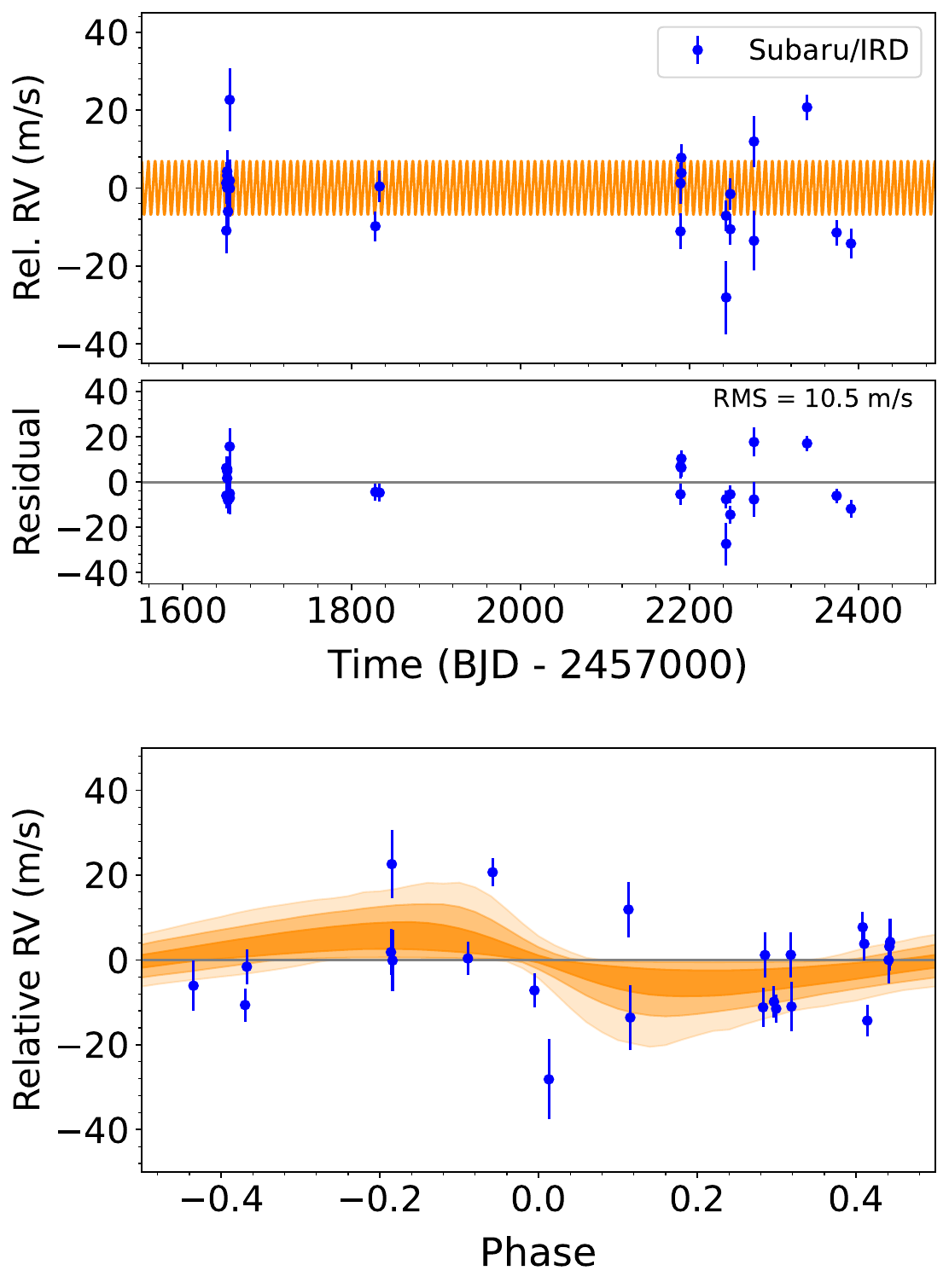}{0.5\textwidth}{(b)}}
	\caption{(a) Detrended and phase-folded light curves of TOI-782b (upper five curves) and their residuals from the best joint-fitting model (lower five curves). The black, blue, green, orange, and red points (from top to bottom in each five data sets) indicate 5 minutes binned data from the TESS, $g$, $r$, $i$, and $z_s$ bands, respectively. Gray lines indicate the best-fit models. The plots are vertically shifted for display. (b) Top: time series of RV data from Subaru/IRD (blue) along with the 1$\sigma$, 2$\sigma$, and 3$\sigma$ credible regions calculated from the posteriors of the joint-fitting analysis (from thick to thin orange). Error bars represent the estimated 1$\sigma$ uncertainties without a jitter term. Middle: residuals of the RV data from the best joint-fitting model. Bottom: same as the top panel, but phase folded with the orbital period of TOI-782\,b. Areas from dark to light orange indicate 1, 2, and 3$\sigma$ credible regions, respectively.
	\label{fig:lc_TOI-782b}
	}
\end{figure*}

\begin{figure*}[ht]
    \gridline{\fig{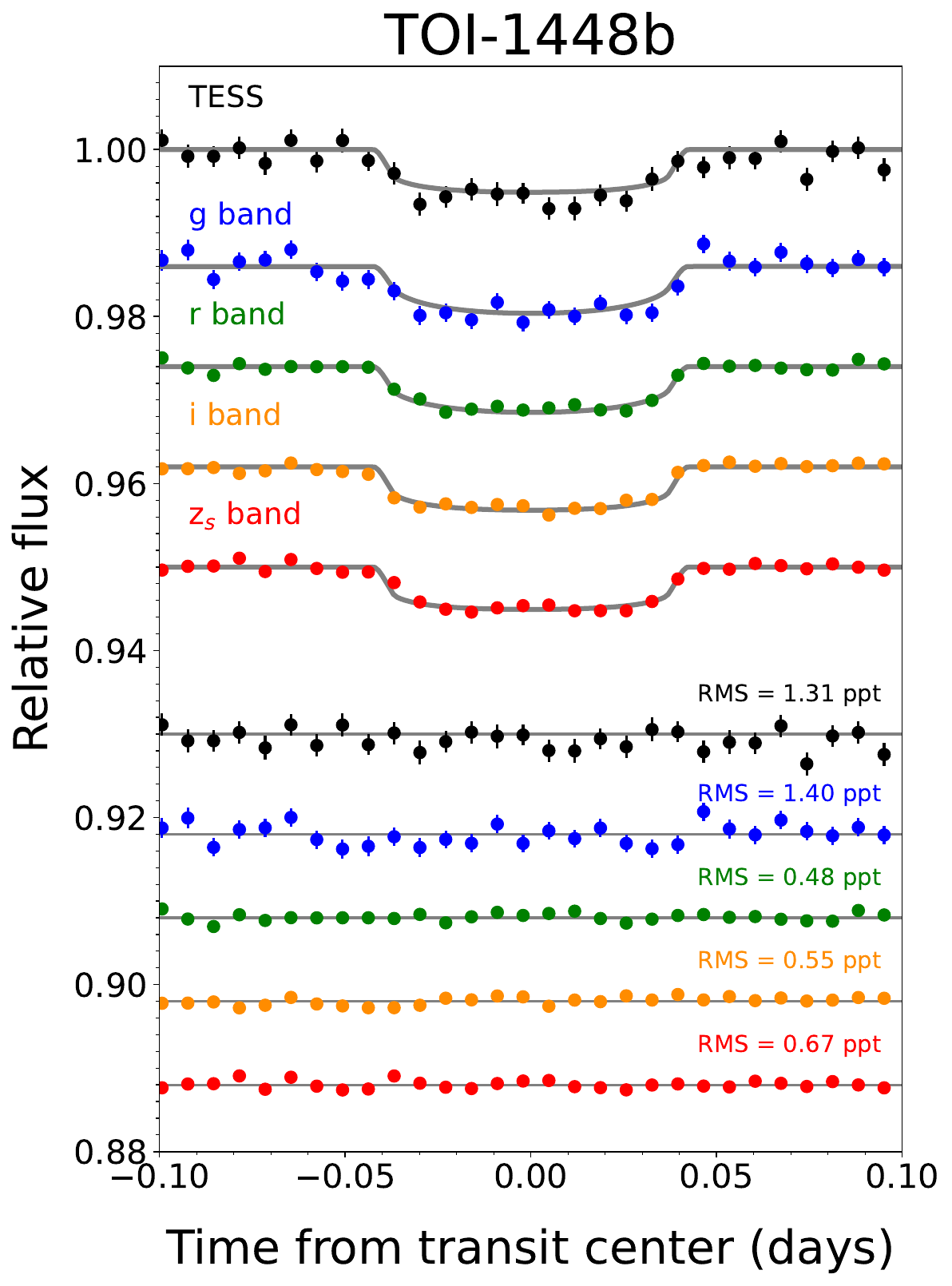}{0.5\textwidth}{(a)}
              \fig{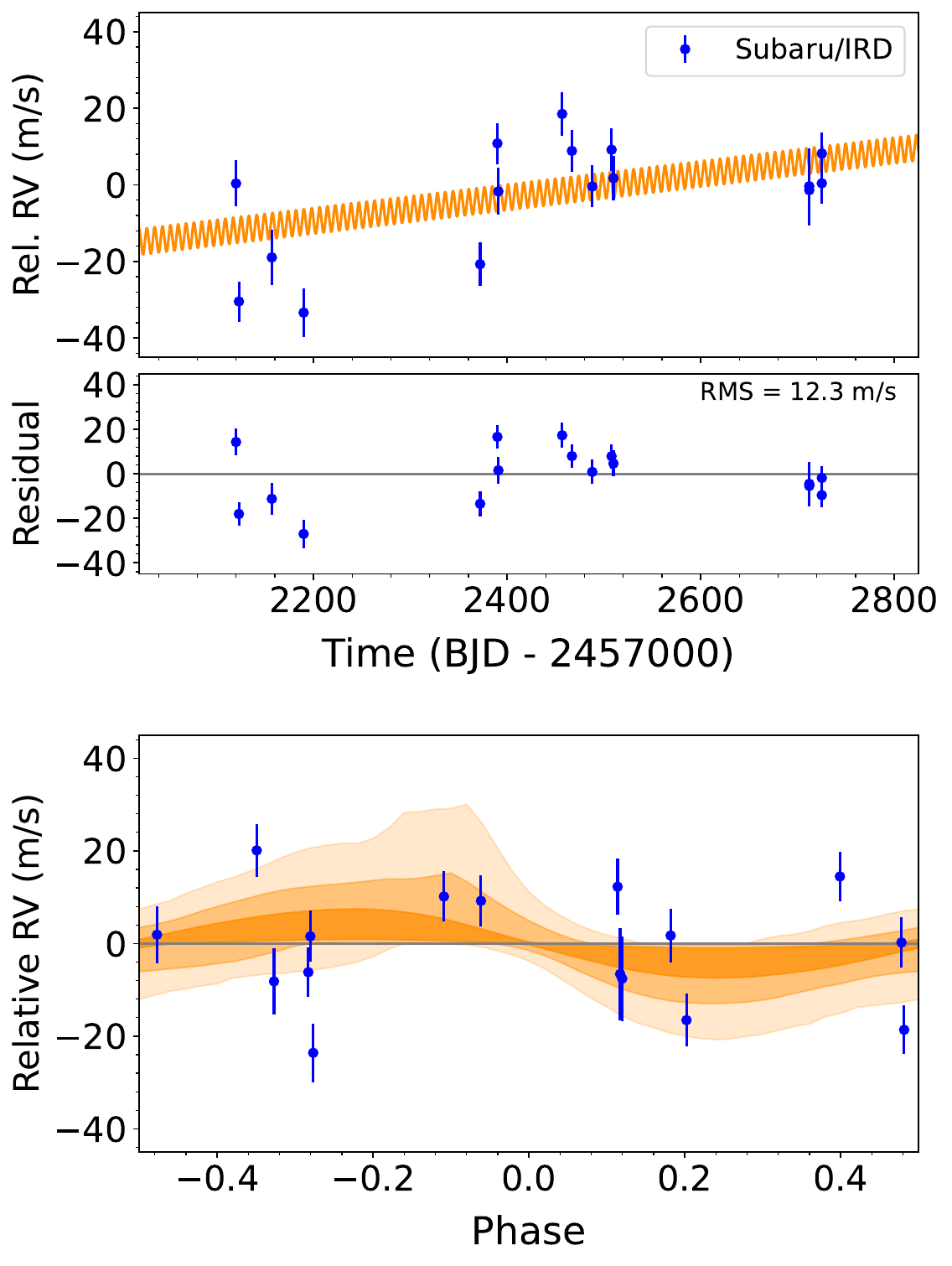}{0.5\textwidth}{(b)}}
	\caption{Same as Figure \ref{fig:lc_TOI-782b} but for TOI-1448\,b.
	\label{fig:lc_TOI-1448b}
	}
\end{figure*}

\begin{figure*}[ht]
    \gridline{\fig{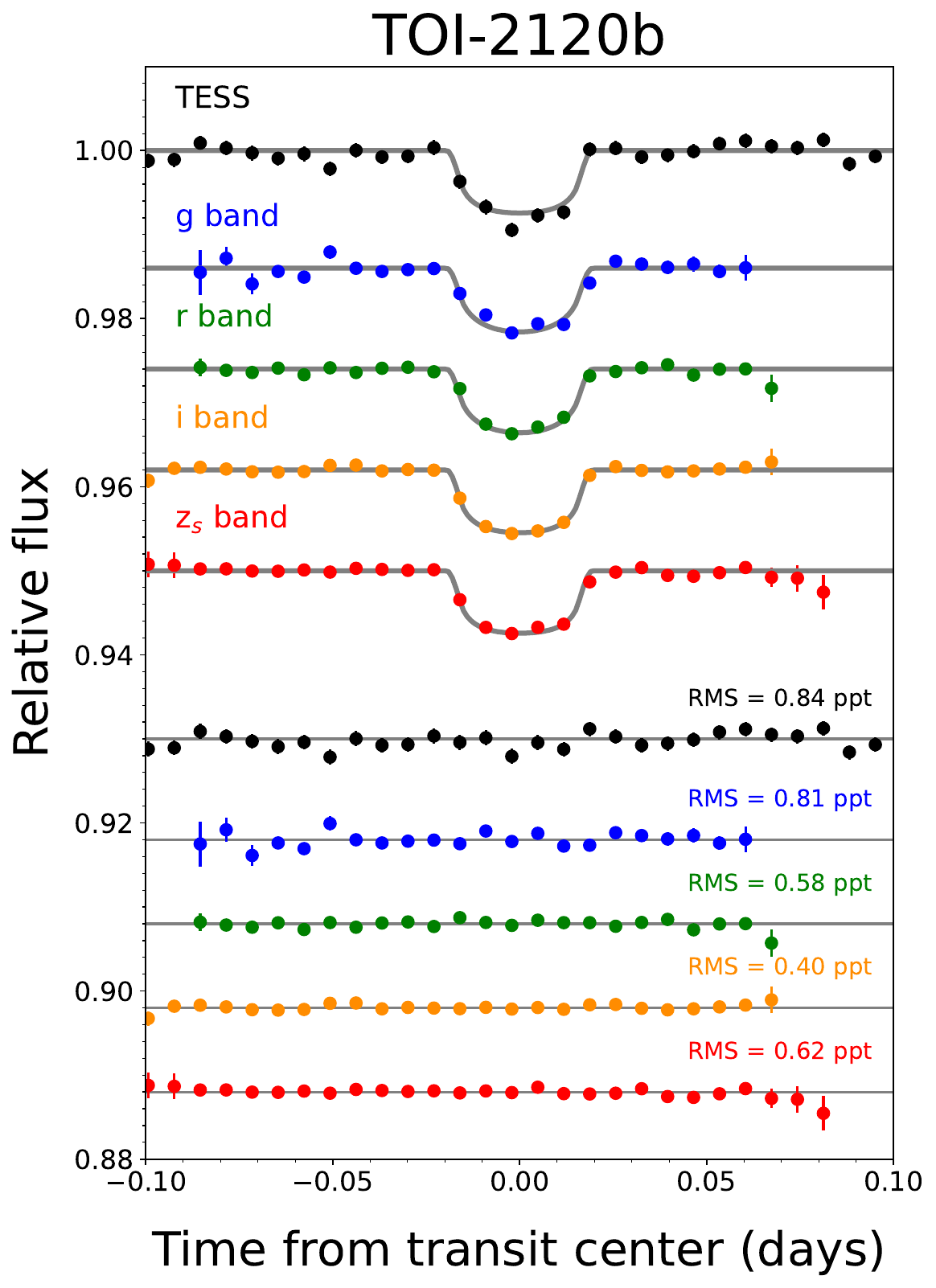}{0.5\textwidth}{(a)}
              \fig{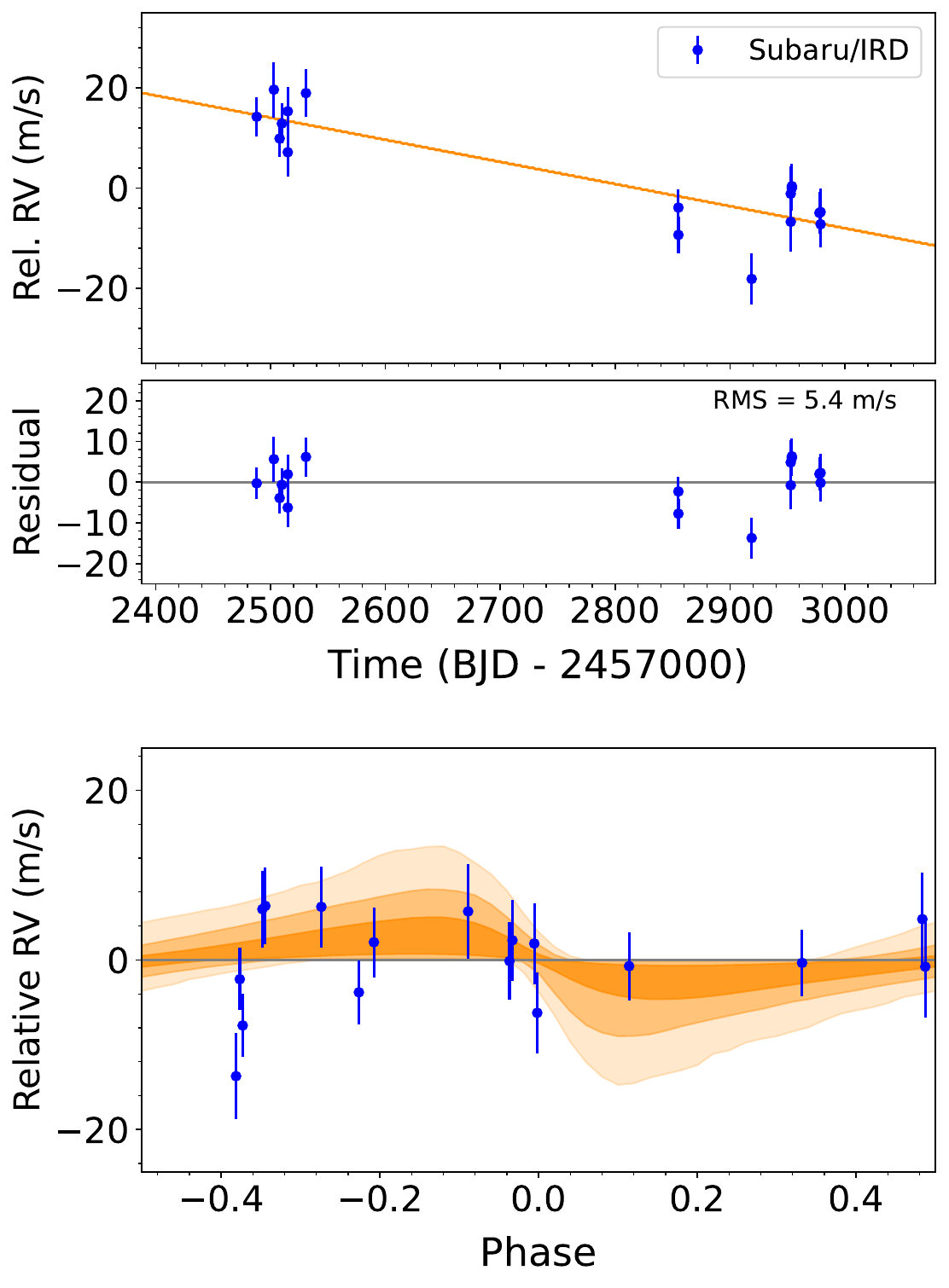}{0.5\textwidth}{(b)}}
	\caption{Same as Figure \ref{fig:lc_TOI-782b} but for TOI-2120\,b. A linear trend in the RV data was injected in the joint analysis of TOI-2120.
	\label{fig:lc_TOI-2120b}
	}
\end{figure*}

\begin{figure*}[ht]
    \gridline{\fig{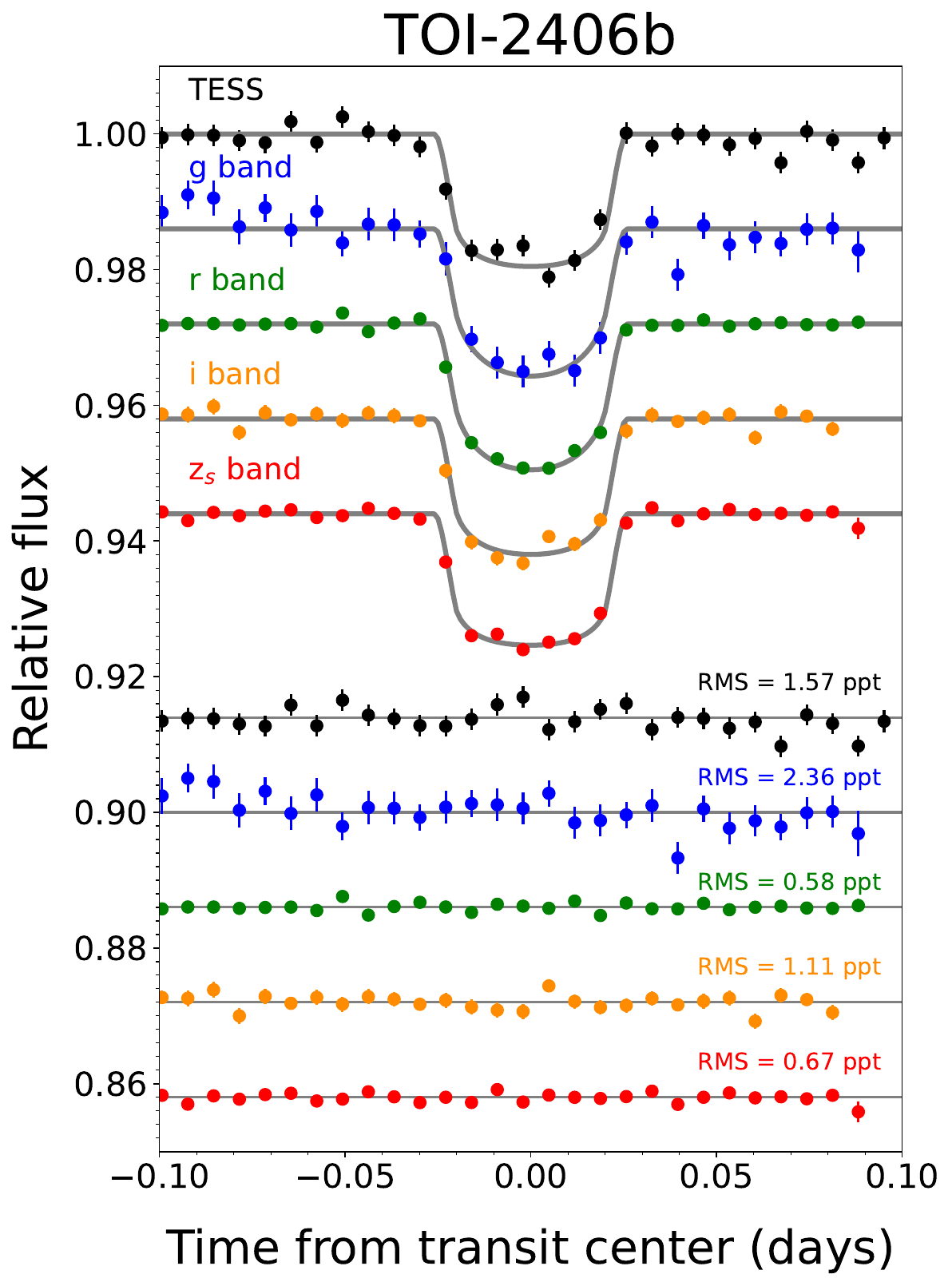}{0.5\textwidth}{(a)}
              \fig{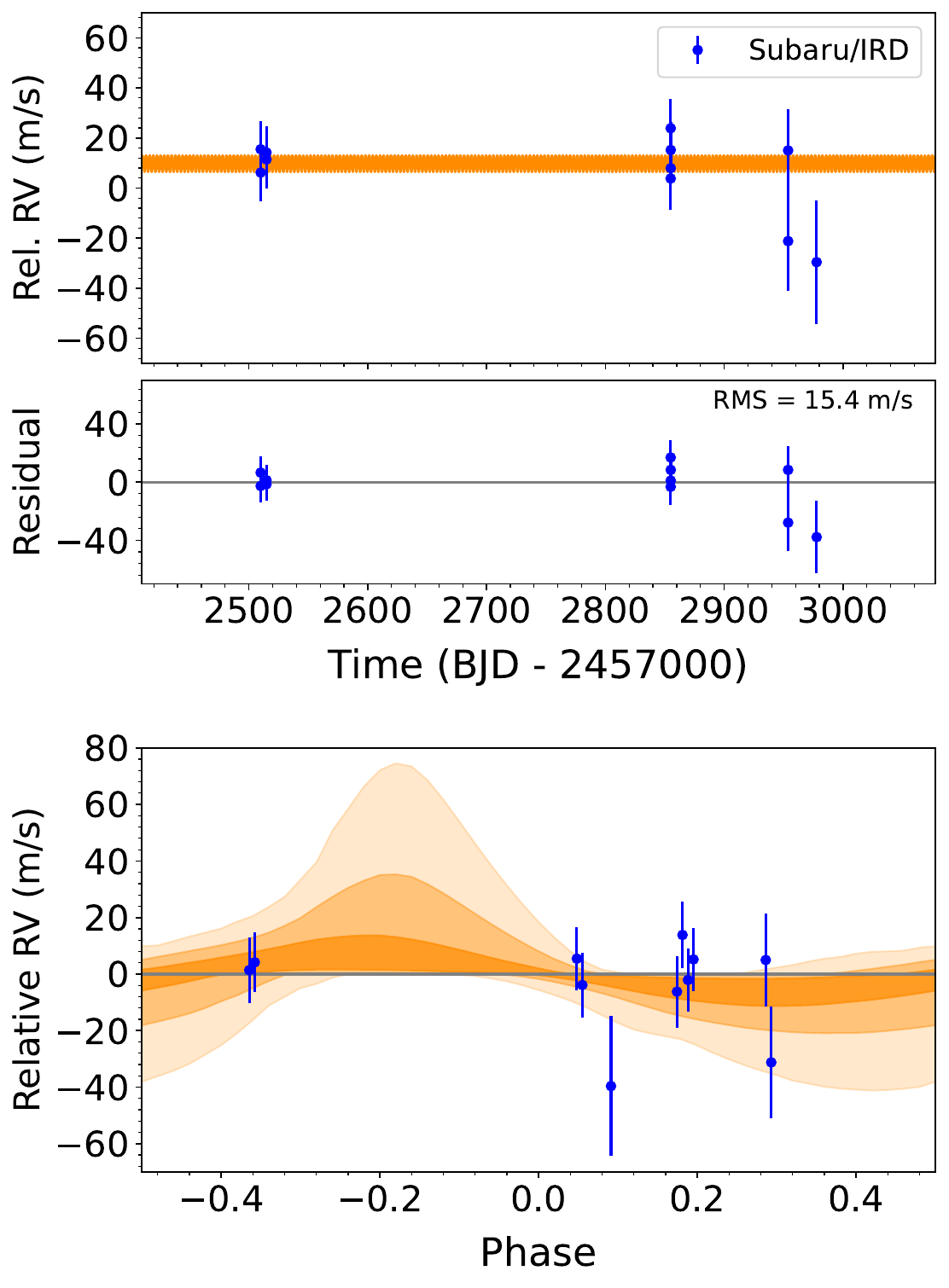}{0.5\textwidth}{(b)}}
	\caption{Same as Figure \ref{fig:lc_TOI-782b} but for TOI-2406\,b.
	\label{fig:lc_TOI-2406b}
	}
\end{figure*}

\subsection{Properties of the Planets}

To derive the physical properties of the planets, we analyzed the TESS light curves, ground-based transit light curves, and RV data of each planetary system in the following steps. First, we modeled the TESS light curves with a transit + Gaussian Process (GP) model to determine the hyperparameters of the GP model.
Next, we modeled the ground-based transit light curves with a transit + GP model using the posteriors of the transit parameters obtained from the above analysis as priors and then determined the hyperparameters of the GP model for the ground-based data.
Finally, we jointly modeled all the light curves and RV data with transit and RV models along with the predetermined hyperparameters of the GP models to derive the final values and uncertainties of the planetary parameters. 
We describe each step in more detail in the following subsections.

\subsubsection{TESS Light Curves}
\label{sec:trfit_TESS}
 
For the transit model, we adopted the Mandel \& Agol model \citep{2002ApJ...580L.171M} implemented by {\tt PyTransit} \citep{Parviainen2015} with the following parameters: scaled semimajor axis $a_R$\,($=a/R_s$, where $a$ is the semimajor axis), impact parameter $b$, planet-to-star radius ratio $k$\,($=R_p/R_s$, where $R_p$ is the planetary radius), eccentricity $e$, the argument of periastron $\omega$, orbital period $P$, individual transit times $T_{{\rm c},i}$, and two coefficients $u_1$ and $u_2$ of a quadratic limb-darkening law. Note that the model is supersampled, i.e., averaged over the exposure time.
At this stage, we set both $e$ and $\omega$ to zero and $P$ to the value provided by the TOI catalog.
For $u_1$ and $u_2$, we applied Gaussian priors with the values and uncertainties of the stellar parameters given by {\tt LDTk} \citep{Parviainen2015b}. Note that we increased the uncertainties of $u_1$ and $u_2$ given by {\tt LDTk} by a factor of 3 to account for the possible systematics in the stellar models.
We assumed uniform priors for the other parameters. 

Simultaneously with the transit model, we also modeled the time-correlated noise in the light curves using a GP model implemented in {\tt celerite} \citep{2017AJ....154..220F} with a stochastically driven, damped simple harmonic oscillator (SHO) kernel function, where the model parameters are the frequency of the undamped oscillation $\omega_0$, the scale factor to the amplitude of the kernel function $S_0$, and the quality factor $\mathcal{Q}$. We set $\mathcal{Q}$ to unity in all sectors to avoid overfitting, while leaving $\omega_0$ and $S_0$ free for each sector to account for the different noise properties from sector to sector. We also fit a white jitter term in the flux for each sector.
We ran an MCMC analysis for each target using {\tt emcee} to estimate the posterior distributions of the parameters.  

\subsubsection{Ground-based Light Curves}
\label{sec:trfit_ground}

We simultaneously modeled the ground-based, multiband transit light curves of each target with a transit + GP model.
We used the same transit model as described in Section \ref{sec:trfit_TESS}, but left $k$ free for each band in order to search for chromaticity in $k$, which can be a sign of flux contamination in the photometric aperture.
The limb-darkening parameters $u_1$ and $u_2$ were fitted with Gaussian priors for each band in the same way as for the TESS light curves.
Note that all the measured fluxes of TOI-2120 are contaminated by a nearby faint background star (TIC\,389900766), as mentioned in Section \ref{sec:validation}. We, therefore, corrected the contamination to TOI-2120\,b by replacing the transit model $\mathcal{F}_{\rm tr}$ with $(\mathcal{F}_{\rm tr} + f_{\rm cont}) / (1 + f_{\rm cont})$, where $f_{\rm cont}$ is the flux ratio of the target to the contaminating source (1.3\%, 1.6\%, 0.78\%, and 0.79\% for  the $g$, $r$, $i$, and $z_s$ bands, respectively, assuming that all the fluxes were contaminated in the photometric apertures).
The time-correlated noise in the ground-based light curves was computed by a GP model as a function of time in the {\tt celerite} package whose covariance function is an approximate Mat\'{e}rn 3/2 kernel with signal standard deviation $\sigma$ and length scale $\rho$. 
The kernel function has the same $\rho$ in all light curves of each transit event but different $\sigma$.
Note that this kernel is a specific version of the SHO kernel (Section \ref{sec:trfit_TESS}) with no oscillation term. We applied this simplified form to the ground-based light curves considering that their observational duratios are shorter than the stellar variability timescale.
The MuSCAT2 data exhibit nonnegligible systematics arising from telescope drifts. Therefore, for the time-correlated noise model of the MuSCAT2 data, we multiplied the GP model by a linear function of $\Delta x$ and $\Delta y$, where $\Delta x$ and $\Delta y$ denote stellar displacements in the detector in the $x$- and $y$-directions, respectively.
We performed an MCMC analysis for each target using {\tt emcee} with uniform priors for all parameters except for $u_1$ and $u_2$.

\subsubsection{Joint analysis of the Radial Velocities and Light Curves}
\label{sec:joint}

From a preliminary analysis of the RV data, we did not detect significant ($> 3\sigma$) planetary signals in any of the systems.
Nonetheless, the RV data are still useful for placing upper bounds on the planetary masses and orbital eccentricities. Because both the RV and transit light-curve models depend on eccentricity, the best constraint on the mass and eccentricity can be obtained by a joint analysis of the RV and transit light-curve data. We therefore simultaneously modeled the RV data and all the available transit light curves (from both TESS and the ground) of each system. 
The parameters of the RV model include RV amplitude $K$, orbital inclination $i$, eccentricity parameters $e$ and $\omega$ (fitted as $\sqrt{e} \sin \omega$ and $\sqrt{e} \cos \omega$), orbital period $P$, time of periastron passage $t_0$, RV offset $v_0$, and RV jitter $\sigma_{v, {\rm jit}}$. We additionally included a parameter for a linear slope in the RV, $\gamma$, for the data of TOI-1448\,b and TOI-2120\,b, for which a preliminary fit of the RV data only indicated that a model with a linear slope was preferred over a model without a slope, determined by corresponding values of the Bayesian information criterion of 3.8 and 21.8, respectively.
The parameters for the transit model are the same as in Section \ref{sec:trfit_ground}, but this time the assumption of a circular orbit was removed. In addition, we assumed a constant orbital period, which introduces a reference transit time $T_{\rm c,0}$ for each system, because we did not detect significant variations in the individual transit timings ($T_{{\rm c},i}$) measured in the TESS and ground-based light curves from a constant period in any system. We also adopted a constant radius ratio $k$ across the bands for each system because we did not detect any significant chromaticity in $k$ in the analyses of Sections \ref{sec:trfit_TESS} and \ref{sec:trfit_ground}. The hyperparameters of the GP for the TESS and ground-based light curves were set to the best-fit values derived in Sections \ref{sec:trfit_TESS} and \ref{sec:trfit_ground}, respectively, to suppress the number of free parameters. Note that the correlations between the hyperparameters and other physical parameters are small enough that fixing the hyperparameters at the best-fit values does not have much effect on the posterior distributions of the hyperparameters.

With the above parameterization, we performed an MCMC analysis using {\tt emcee}.
Note that in this analysis $i$ and $t_0$ are not fitted but $i$ is converted from $b$, $a_R$, $e$, and $\omega$ by the equation of $b=a_R \cos i (1-e^2)/(1+e\sin \omega)$ and $t_0$ is converted from $T_{\rm c, 0}$, $P$ and Kepler's equation. In addition, we use log stellar density $\log \rho_s$ as the fitting parameter instead of $a_R$, which is converted from $\rho_s$ and $P$ by the equation $a_R = (\rho_s GP^2 / 3\pi)^{1/3}$ (assuming $k^3 \ll 1$), where $G$ is the gravitational constant.

For the priors, we applied a two-sided Gaussian distribution for $\rho_s$ using the values derived in Section~\ref{sec:photometric_properties}. We also applied the joint probability distribution for $e$ and $\omega$ of \citet{2014MNRAS.444.2263K}, Equation (23) in that paper), which is a conditional probability given that the planet has a transit geometry adopting a beta function as the underlying prior for $e$. For the two coefficients of the beta function, we adopted $\alpha=1.58$ and $\beta=4.4$ from \citet{2019AJ....157...61V}, which were derived from a sample of Kepler single-transiting systems. Uniform priors were applied to all the other parameters.

The derived posterior values (median and 1$\sigma$ boundaries or 2$\sigma$ upper limits) are listed in Table~\ref{tab:joint}. The transit light curves detrended and phase folded by the best-fit parameters, and the time series and phase-folded RV data along with the posterior RV models,  for TOI-782b, TOI-1448b, TOI-2120b, and TOI-2406b are shown in Figures~\ref{fig:lc_TOI-782b}, \ref{fig:lc_TOI-1448b}, \ref{fig:lc_TOI-2120b}, and \ref{fig:lc_TOI-2406b}, respectively.

\begin{deluxetable*}{lcccccr}
\tablenum{5}
\tablecaption{Posterior Values of the Parameters Used in the Joint Fitting}
\tablewidth{0pt}
\tablehead{
\colhead{Parameters} &  \colhead{Unit} & \colhead{}  & \multicolumn2c{Values}  & \colhead{}\\
\colhead{} & \colhead{} & \colhead{TOI-782\,b} & \colhead{TOI-1448\,b} & \colhead{TOI-2120\,b} & \colhead{TOI-2406\,b}
}
\startdata
$P$ &  days  &  $8.0240015\,^{+0.0000077}_{-0.0000072}$ & $8.112245 \pm 0.000018$ & $5.7998164\pm0.0000035$ & $3.0766891\pm0.0000024$ \\
$T_{c,0}$ & BTJD \tablenotemark{a} & $1577.04189\,^{+0.00085}_{-0.00093}$ & $1713.3375\pm0.0015$ & $1795.82368\pm0.00041$ & $2115.97600\,^{+0.00031}_{-0.00029}$ \\
$b$ & & $0.46\,^{+0.12}_{-0.25}$ & $0.21\,^{+0.16}_{-0.14}$ & $0.55\,^{+0.07}_{-0.10}$ & $0.09\,^{+0.09}_{-0.06}$ \\
$k$ & & $0.0606\pm0.0015$ & $0.0663 \pm 0.0011$ & $0.0796\pm0.0014$ & $0.1287\pm0.0013$ \\
$\rho_s$ & g cm$^{-3}$ & $7.99\pm0.54$ & $9.51\pm0.55$ & $20.6\,^{+1.7}_{-1.5}$ & $27.4\pm1.9$ \\
$\sqrt{e}\sin{\omega}$ & & $0.31\,^{+0.14}_{-0.19}$ & $-0.03\,^{+0.13}_{-0.15}$ & $0.48\,^{+0.08}_{-0.10}$ & $-0.31\,^{+0.05}_{-0.04}$ \\
$\sqrt{e}\cos{\omega}$ & & $0.05\pm0.33$ & $-0.01\,^{+0.30}_{-0.27}$ & $0.07\,^{+0.33}_{-0.35}$ & $0.10\,^{+0.30}_{-0.36}$ \\
$K$ & m s$^{-1}$ & $5.7\,^{+3.3}_{-3.0}$ ($<$\,11.6) & $<$\,12.5 & $<$\,7.3 & $<$\,23.7 \\
$\gamma$ & m s$^{-1}$ d$^{-1}$ & --- & $0.041 \pm 0.019$ & $-0.045\pm0.007$ & ---\\ 
$\sigma_{\rm rv, jitter}$ & m s$^{-1}$ & $9.0\,^{+2.1}_{-1.7}$ & $13.0\,^{+4.0}_{-2.9}$ & $3.5\pm2.0$ & $4.7\,^{+5.9}_{-3.3}$\\
\enddata
\label{tab:joint}
\tablecomments{The reported values, uncertainties, and upper limits represent the median, 1$\sigma$ lower and upper boundaries, and 2$\sigma$ upper boundaries of the posterior probability distributions, respectively.}
\tablenotetext{a}{Barycentric TESS Julian Date, which corresponds to Barycentric Julian Date (BJD) - 2457000.}
\end{deluxetable*}

\subsubsection{Results}

Combining the results of the joint analysis in Section \ref{sec:joint} with the stellar properties derived in Section \ref{sec:photometric_properties},  we derive the physical properties of the planets as listed in Table \ref{tab:joint}. We find that TOI-782\,b, TOI-1448\,b, TOI-2120\,b, and TOI-2406\,b have radii of $2.734\pm0.085$\,$R_\oplus$, $2.749 ^{+0.067}_{-0.063}$\,$R_\oplus$, $2.122\pm 0.065$\,$R_\oplus$, and $2.860 ^{+0.063}_{-0.069}$\,$R_\oplus$, respectively, all of which are within the sub-Neptune regime. Our value for TOI-2406\,b is consistent with that measured by \cite{2021A&A...653A..97W} within the uncertainties but is more precise thanks to the additional TESS data taken after their work and our high-precision multiband transit light curves taken with MuSCAT3. 

\begin{figure*}[ht]
    \gridline{\fig{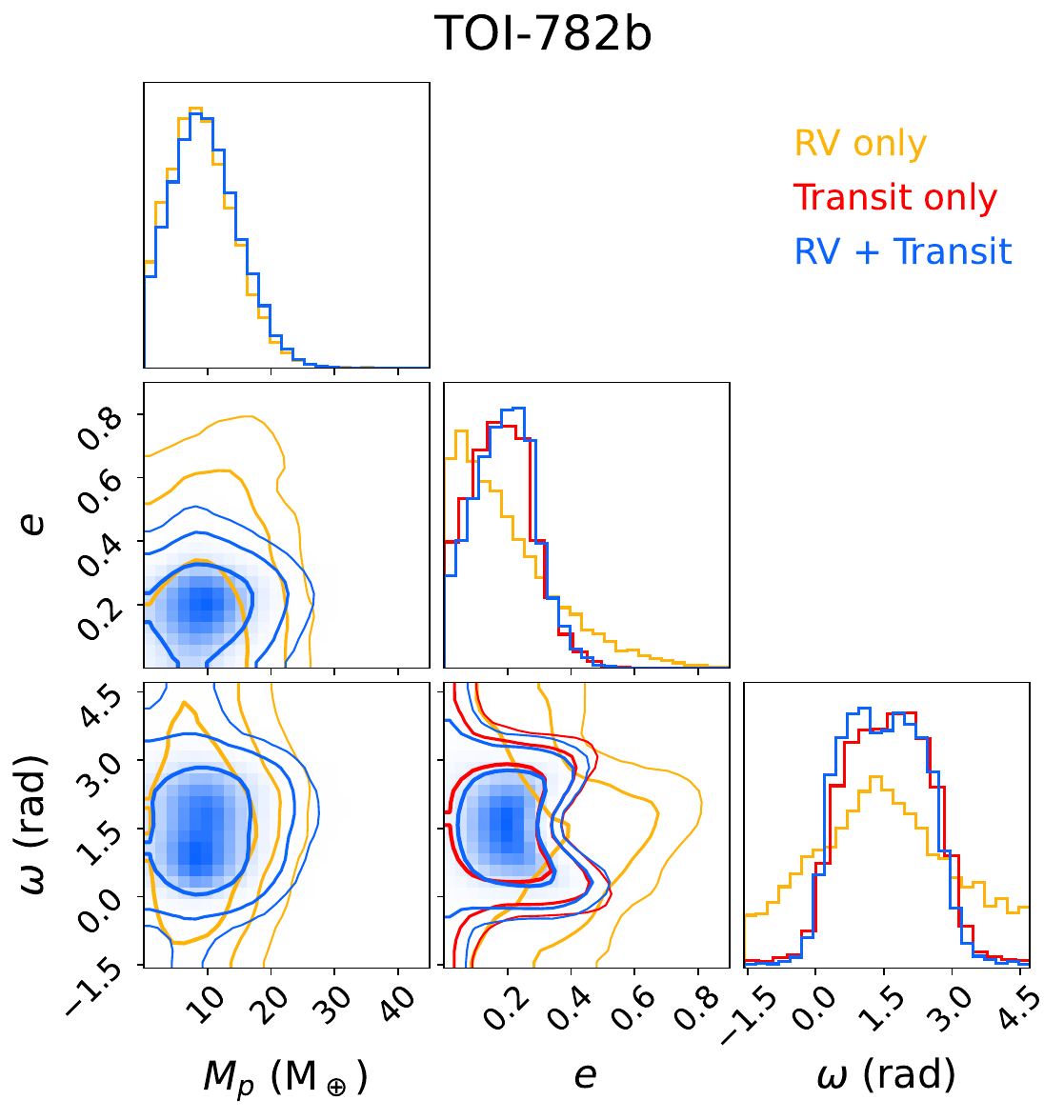}{0.4\textwidth}{(a)}
            \fig{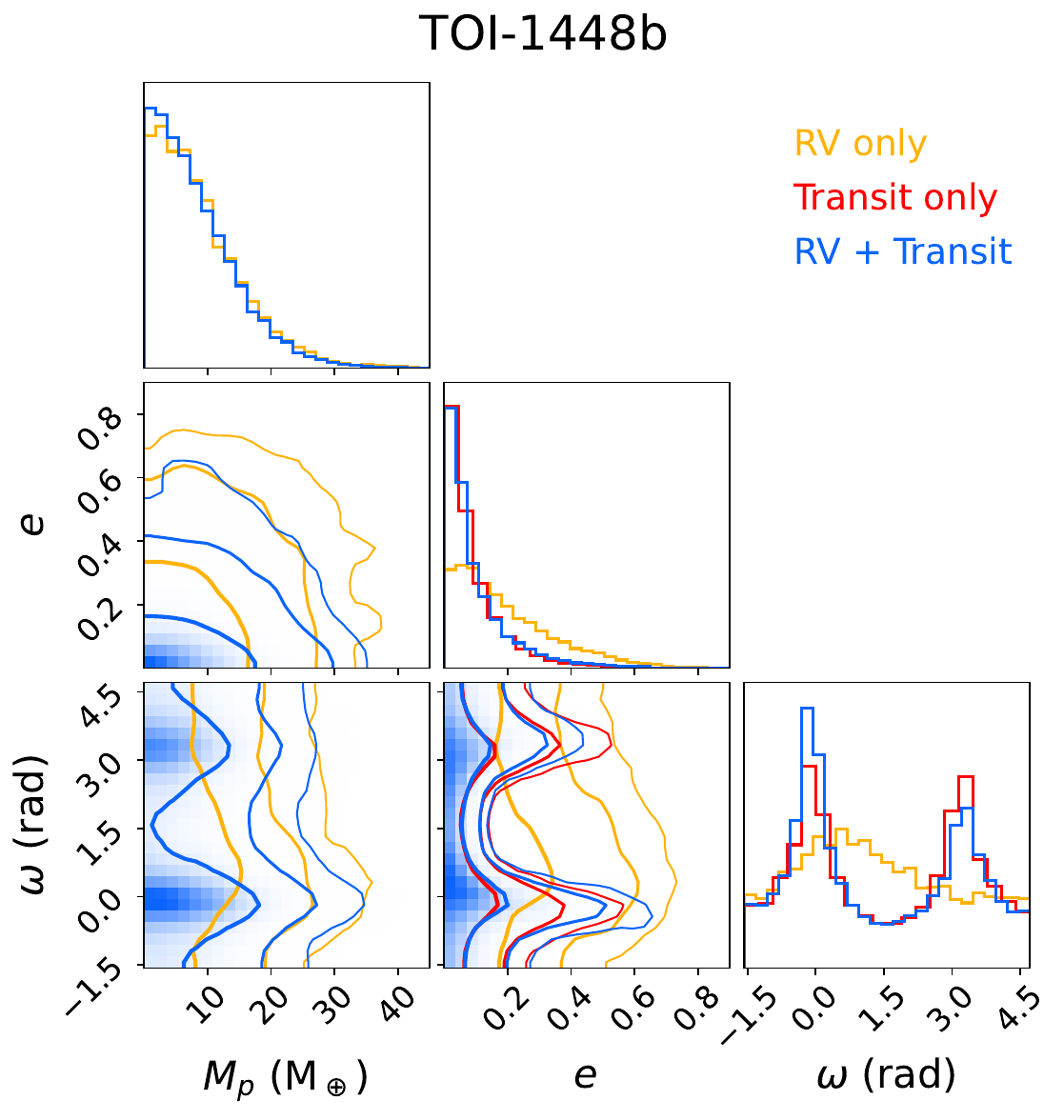}{0.4\textwidth}{(b)}}
    \gridline{\fig{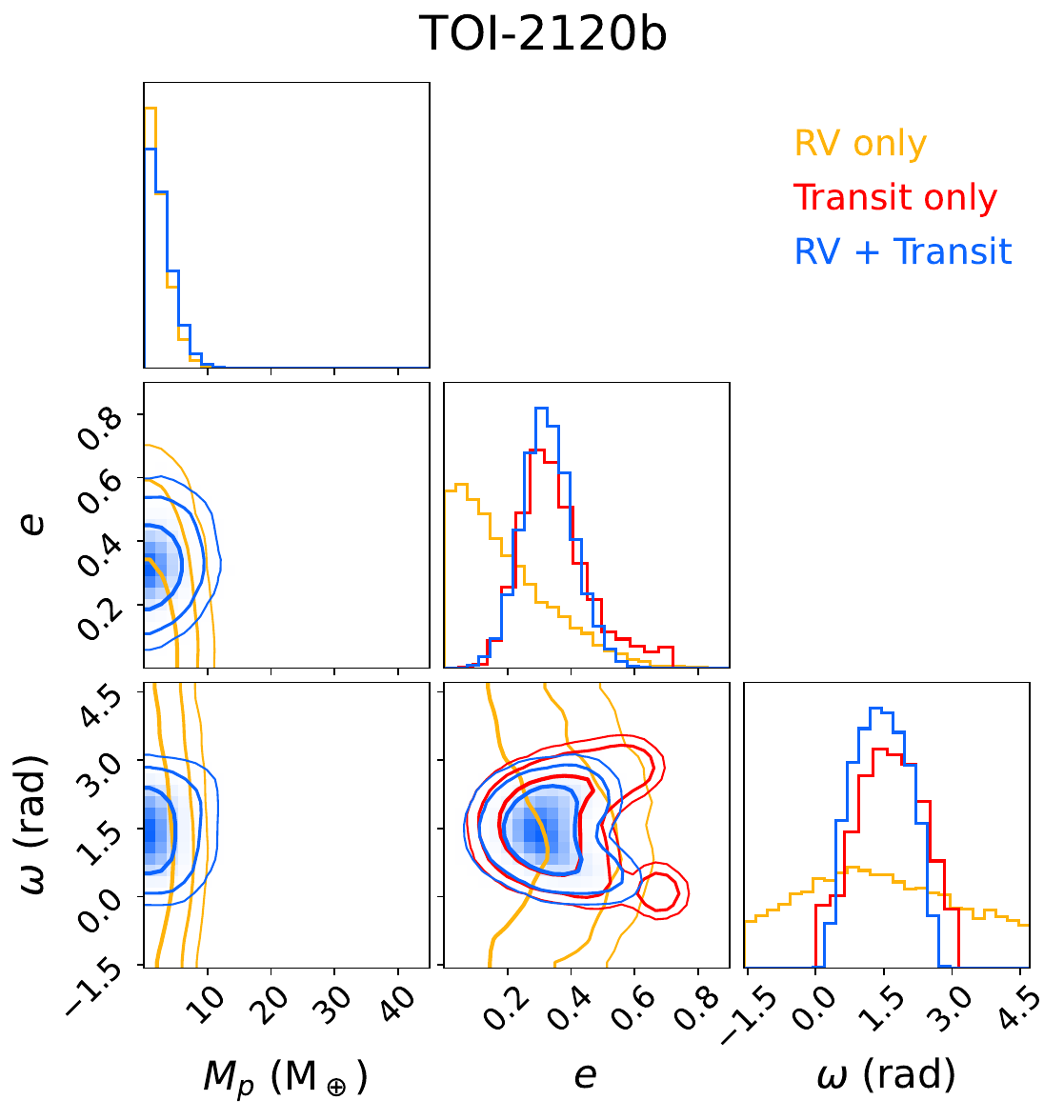}{0.4\textwidth}{(c)}
            \fig{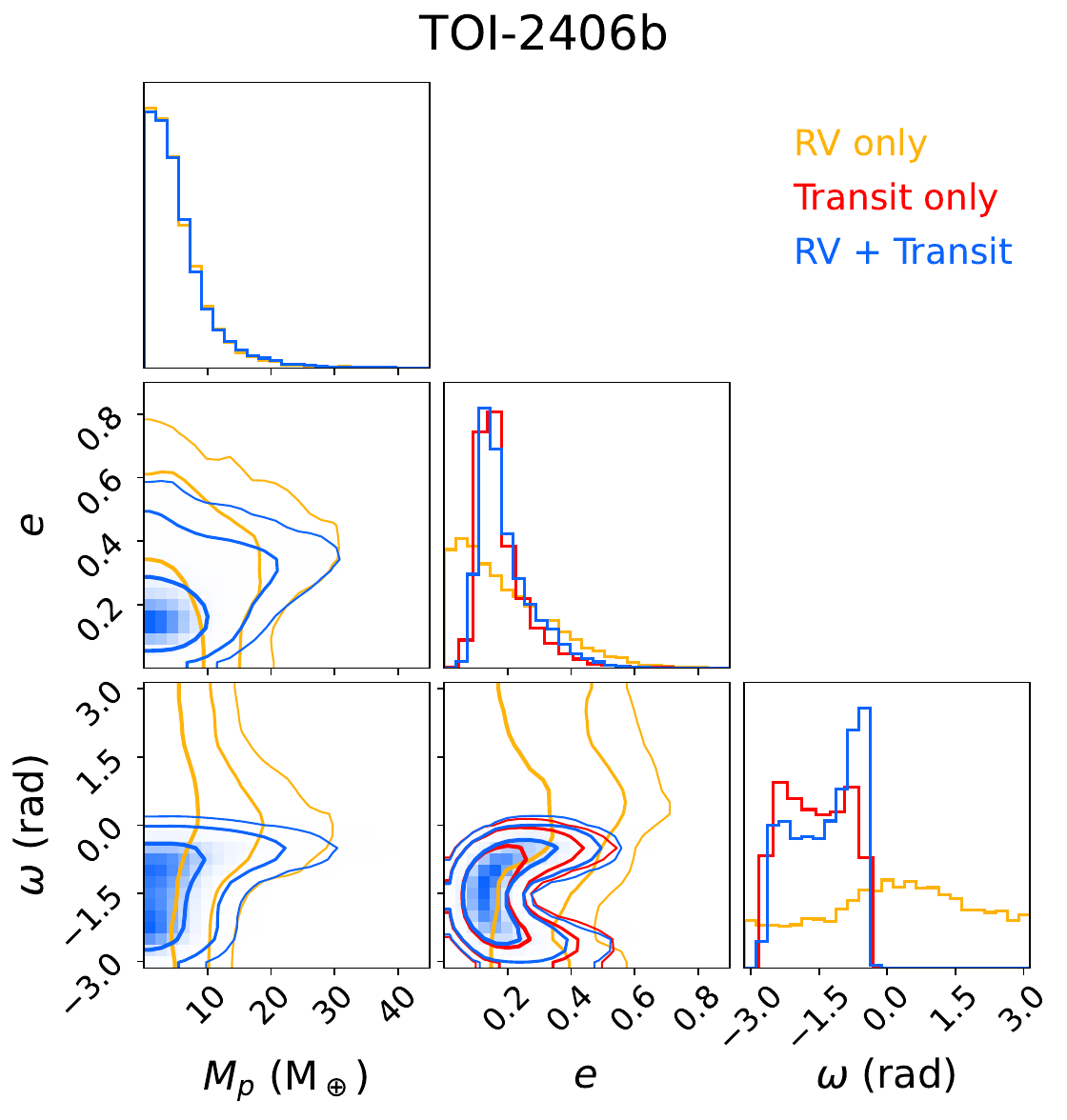}{0.4\textwidth}{(d)}}
	\caption{(a) Corner plots for the posterior samples of $M_p$, $e$, and $\omega$ for TOI-782\,b. Yellow, red, and blue contours indicate the posteriors from the RV-only fit, the transit-only fit, and the RV+transit joint fit, respectively. Contours with thick to thin solid lines indicate 68\%, 95\%, and 99\% confidence regions, respectively. The density of the posterior samples from the joint analysis is indicated by the color map. (b), (c), (d) Same as (a) but for TOI-1448\,b, TOI-2120\,b, and TOI-2406\,b, respectively.
	\label{fig:corner}
	}
\end{figure*}

Two-dimensional posterior distributions between $M_p$, $e$, and $\omega$ obtained from the joint analysis are shown in blue in Figure \ref{fig:corner}.
Although we are unable to measure the masses of any of the planets with a high enough significance, we place 2$\sigma$ upper limits on the masses of 19.1, 19.5, 6.8, and 15.6 M$_\oplus$ for TOI-782\,b, TOI-1448\,b, TOI-2120\,b, and TOI-2406\,b, respectively.
We find tentative linear trends in the RV data of TOI-1448 and TOI-2120 with $\gamma=$ $0.041\pm0.019$ and $-0.044\pm0.007$\,m\,s$^{-1}$\,day$^{-1}$, respectively, which could be due to an additional outer planet or companion, or some unrecognized systematics. Further RV observations are needed to determine the definitive masses of these planets and to confirm or refute the presence of outer planets/companions to TOI-1448\,b and TOI-2120\,b.

While we cannot unambiguously determine the eccentricity due to the lack of significant planetary signals in the RV data, our joint analyses of RV data and transit light curves with stellar density priors show that three of the four planets, TOI-782\,b, TOI-2120\,b, and TOI-2406\,b, have nonzero eccentricity with significances of $\sim$2--4$\sigma$ ($0.19 ^{+0.09}_{-0.11}$, $0.32\pm0.08$, and $0.17 ^{+0.11}_{-0.05}$, respectively). These eccentricities are mostly constrained by the transit-light curves with the stellar density priors. To illustrate this, in Figure~\ref{fig:corner} we overplot the posterior distributions of $M_p$, $e$, and $\omega$ obtained from analyses of the RV data alone (where $T_{c,0}$, $P$ and $i$ are fixed to the values measured in the transit light curves) in yellow and the posterior distributions of $e$ and $\omega$ from analyses of the transit light curve data alone (both TESS and ground-based ones) in red. We note that our conclusion that three of the four planets are likely to have nonzero eccentricity does not depend on the prior on the eccentricity; we also detect nonzero eccentricities for TOI-782\,b, TOI-2120\,b, and TOI-2406\,b at similar significances even with a uniform prior on $e$. We further note that a nonzero eccentricity of TOI-2406\,b was originally found by \cite{2021A&A...653A..97W}, with a uniform prior on $e$, and we confirmed it with almost independent transit data (except for TESS Sectors 3 and 30) and new constraints from the RVs. 

\begin{deluxetable*}{lccccccr}
\tablenum{6}
\tablecaption{Properties of the Four Sub-Neptunes around M dwarfs}
\tablewidth{0pt}
\tablehead{
\colhead{Parameters} & \colhead{Unit} &  \multicolumn5c{Values}\\
\colhead{} & \colhead{} & \colhead{TOI-782\,b} & \colhead{TOI-1448\,b} & \colhead{TOI-2120\,b} & \multicolumn2c{TOI-2406\,b}  & \colhead{} \\
\colhead{} &  \colhead{} & \colhead{}  & \colhead{}  & \colhead{} &  \colhead{This work}  &   \colhead{Wells et al.(2021)}
}
\startdata
Radius & $R_\oplus$  & $2.734\pm0.085$ & $2.749\,^{+0.067}_{-0.063}$ & $2.122\pm0.065$ & $2.860\,^{+0.063}_{-0.069}$ & $2.94^{+0.17}_{-0.16}$ \\
Mass & $M_\oplus$  & $<$\,19.1 & $<$\,19.5 & $<$\,6.8 & $<$\,15.6  & --- \\
Eccentricity &  &  $0.19\,^{+0.09}_{-0.11}$ & $<$\,0.36 & $0.32\pm0.08$ & $0.17\,^{+0.11}_{-0.05}$  & $0.26^{+0.27}_{-0.12}$ \\
Semi-major axis & au &  $0.0578\pm0.0018$ & $0.0567\pm0.0015$ & $0.0377\pm0.0013$ & $0.02267\pm0.00071$  & $0.0228\pm0.0016$ \\
Scaled semi-major axis ($a/R_s$) & & $30.07\pm0.68$ & $32.10\pm0.62$ & $33.21\,^{+0.87}_{-0.83}$ & $23.95\pm0.54$ & $24.0\,^{+1.0}_{-1.1}$  \\
Orbital inclination & deg &  $88.98\,^{+0.51}_{-0.17}$ & $89.62\pm0.26$ & $88.67\,^{+0.14}_{-0.11}$ & $89.80\,^{+0.14}_{-0.19}$ & $89.63^{+0.27}_{-0.35}$ \\
Equilibrium temp. ($A_{\rm Bond}=0.0$) & K & $435\pm6$ & $426\pm6$ & $384\pm6$ & $449\pm7$ & $447\pm15$ \\
Equilibrium temp. ($A_{\rm Bond}=0.3$) & K & $398\pm6$ & $389\pm5$ & $351\pm6$ & $410\pm6$ & --- \\
Insolation & $F_\oplus$ & $5.91\,^{+0.35}_{-0.33}$ & $5.45\,^{+0.29}_{-0.27}$ & $3.61\,^{+0.24}_{-0.22}$ & $6.69\,^{+0.41}_{-0.36}$ & $6.55\,^{+0.94}_{-0.80}$\\
\enddata
\label{tab:planet}
\tablecomments{The reported values, uncertainties, and upper limits represent the median, 1$\sigma$ lower and upper boundaries, and 2$\sigma$ upper limits of the posterior probability distributions, respectively.}
\end{deluxetable*}

\section{Discussion}\label{sec:dis}

In this study, we discovered and followed up four single sub-Neptunes transiting nearby ($<$100\,pc) M dwarfs. Although these four planets are not outliers in the radius-period diagram among known close-in sub-Neptunes around M dwarfs (see Figure \ref{fig:R-P}), we found that at least three of them are likely to have eccentric orbits with $e \sim 0.2-0.3$, which are slightly larger than the majority of known close-in sub-Neptunes with orbital periods of $\lesssim 10$\,days (see Figure \ref{fig:R-e}). One possible explanation for the nonzero eccentricities of {\bf these} planets is that there is an unseen (long-period) planet \citep[e.g.][]{2015PNAS..112...20L} or a substellar companion \citep[e.g.][]{2016ApJ...827....8N} in each system.
Once the disk gas has dissipated, the eccentricity of a planet can be excited by other planets and (sub-)stellar companions through gravitational perturbations, such as planet-planet scattering \citep[e.g.][]{1996Natur.384..619W} and the von Zeipel-Lidov-Kozai mechanism \citep[e.g.][]{2007ApJ...669.1298F}. 
In fact, TOI-1448\,b and TOI-2120\,b may have an unseen siblings as suggested by the linear trend in the RV data (see Figure\,\ref{fig:lc_TOI-2120b}).

Another dynamical process that influences the eccentricity of a close-in planet is tidal interaction between the planet and its host star. As an eccentric planet approaches to its host star, tidal forces circularize its orbit.
Nonzero eccentricities of the close-in sub-Neptunes may suggest that tidal circularization of the planet's orbit is not achieved because the host star is too young or because inefficient tidal dissipation occurs in the planetary interior. 
The estimated ages of the four M dwarfs are all older than $1$\,Gyr (see Section 3.2).
Therefore, all four sub-Neptunes should experience tidal forces from their host stars over 1\,Gyr. The nonzero eccentricities of at least three of the sub-Neptunes are suggestive of less efficient tidal dissipation in their interiors.

\begin{figure*}[ht!]
	\epsscale{1.2}
	\plotone{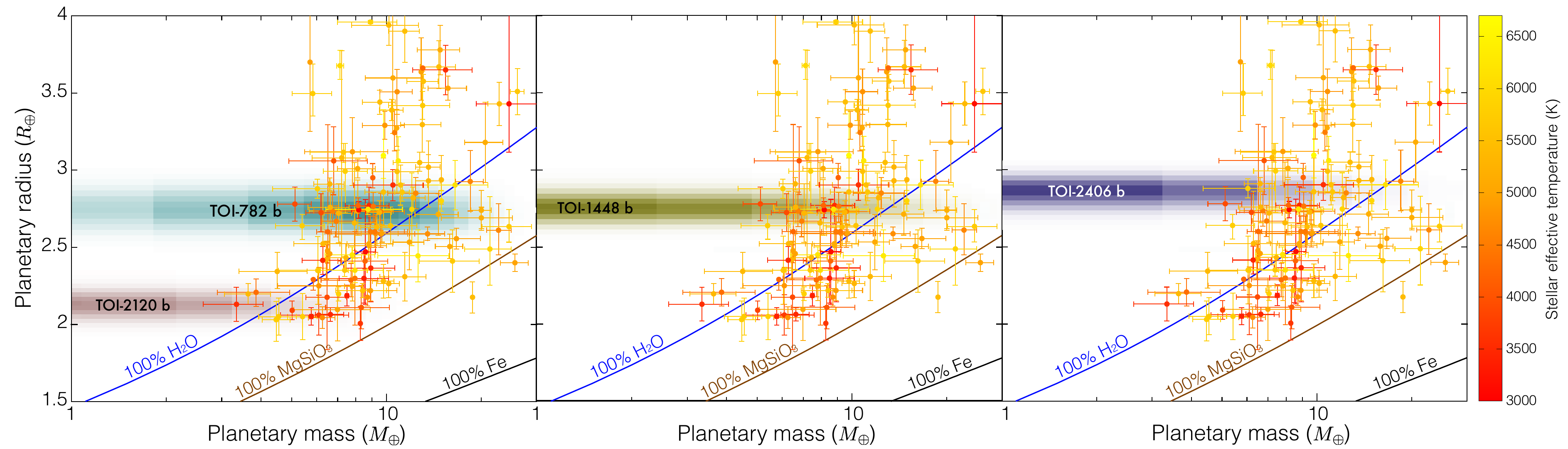}
	\caption{Mass-radius relationship of planets with measured radii of $2-4R_\oplus$ and orbital periods of $\lesssim 50$\,days around Sun-like stars and M dwarfs (data from \url{http://www.exoplanet.eu}). The curves show the interior models calculated for planets composed of pure H$_2$O, MgSiO$_3$, and Fe. We adopted the third-order Birch--Murnagham equation of state (EOS) for MgSiO$_3$ perovskite \citep{2000PhRvB..6214750K,2007ApJ...669.1279S}, the Vinet EOS for $\epsilon$--Fe \citep{2001GeoRL..28..399A}, the Thomas--Fermi--Dirac EOS \citep{1967PhRv..158..876S} at high pressure \citep{2007ApJ...669.1279S,2013PASP..125..227Z}, and AQUA EOS for H$_2$O \citep{2020A&A...643A.105H}.
    We also plotted the contours of mass ranges derived from joint analyses of the four sub-Neptunes validated in this study.
	\label{fig:MR}
	}
\end{figure*}

\begin{figure}[ht!]
	\epsscale{1.1}
	\plotone{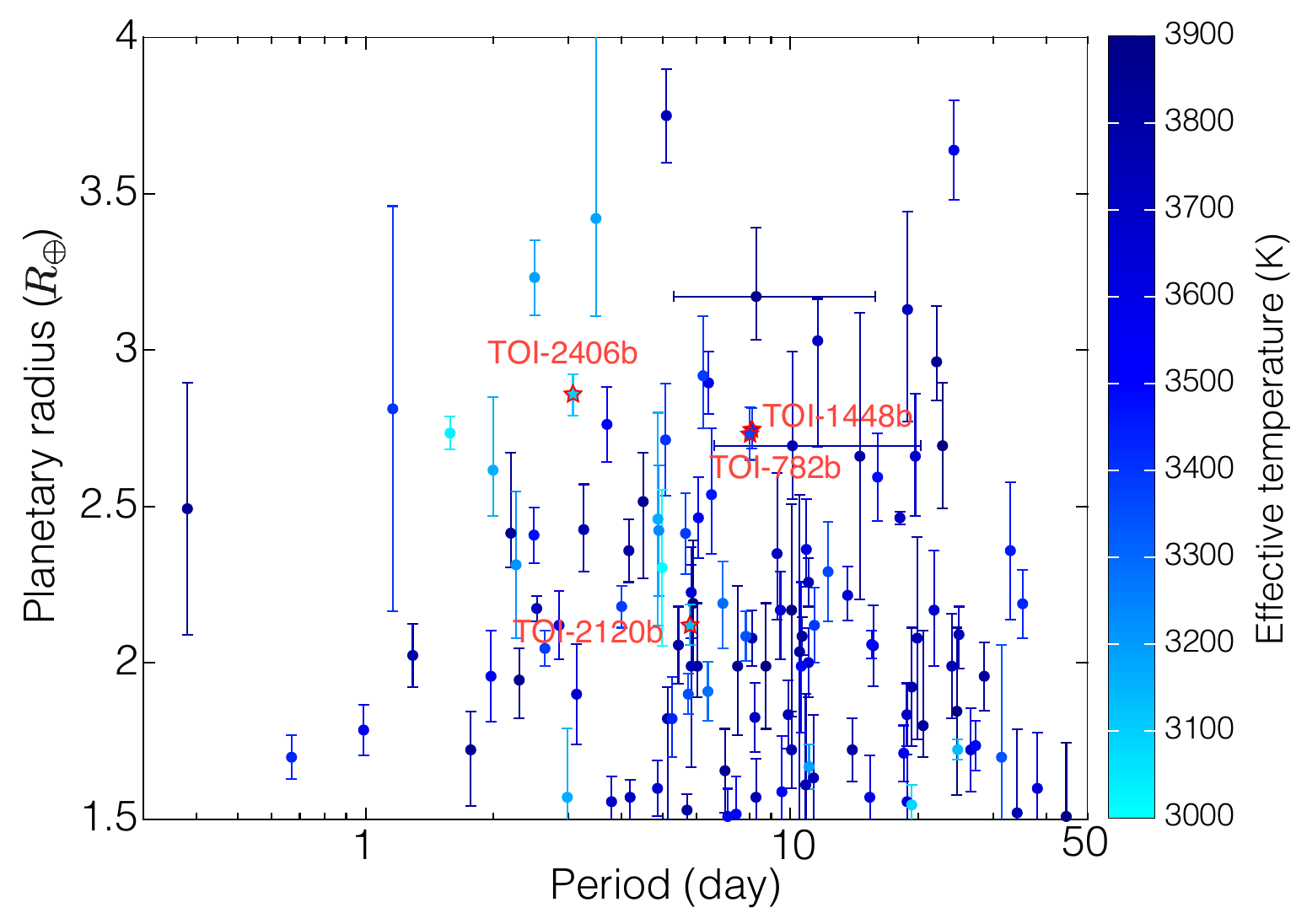}
	\caption{Radius-period relationship of planets with measured radii of $1.5-4R_\oplus$ and orbital periods of $\lesssim 50$\,days around M dwarfs with $T_\mathrm{eff} = 3000-3900$\,K (data from \url{http://www.exoplanet.eu}). The coloured contour represents the effective temperature of a planet-hosting M dwarf. The four close-in, sub-Neptunes validated in this study are shown as red stars.
	\label{fig:R-P}
	}
\end{figure}

\begin{figure}[ht!]
	\epsscale{1.1}
	\plotone{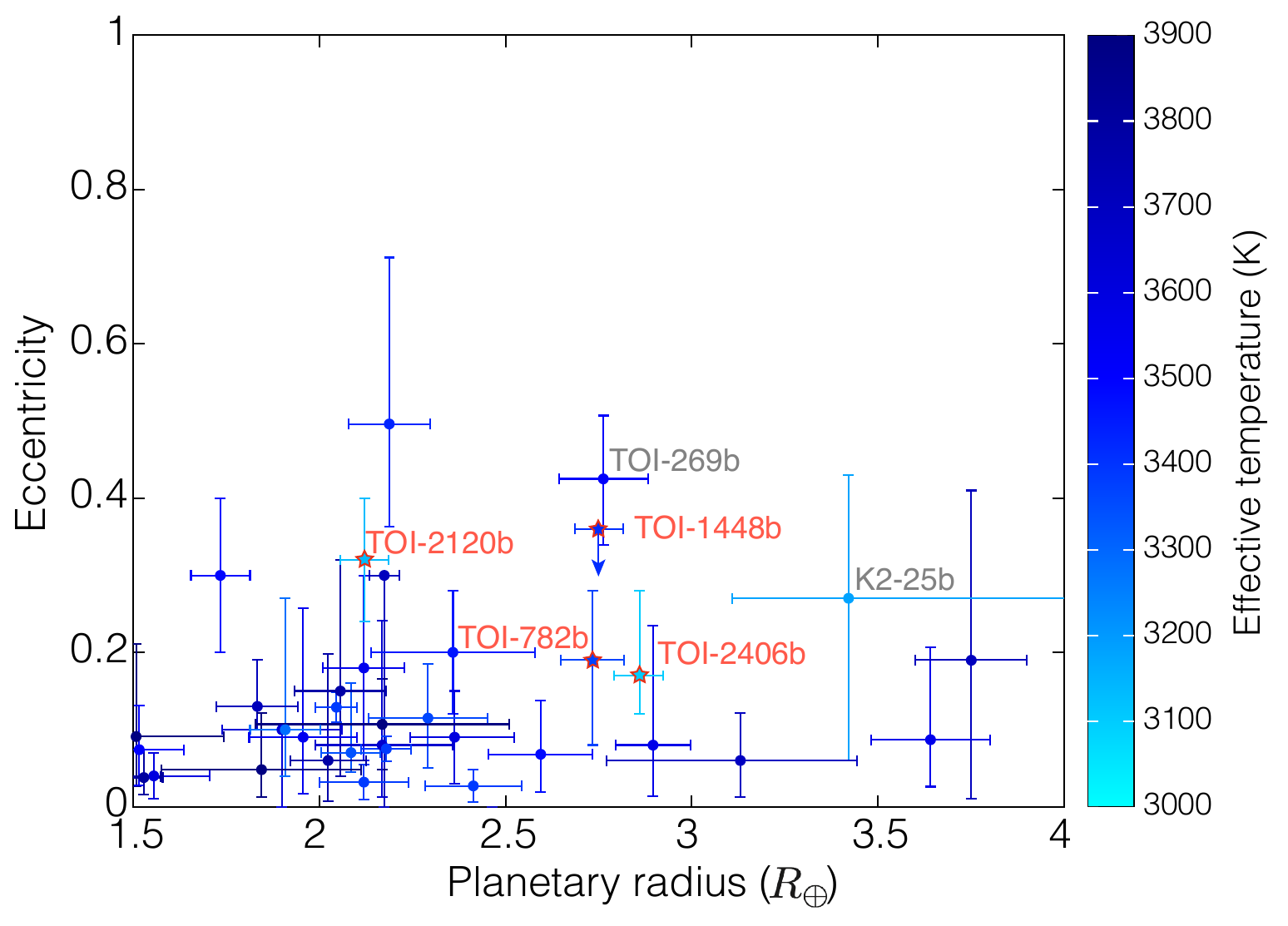}
	\caption{Eccentricity distribution of planets with measured radii of $1.5-4R_\oplus$ and orbital periods of $\lesssim 50$\,days around M dwarfs with $T_\mathrm{eff} = 3000-3900$\,K. Only planets with measured eccentricities are shown. The color contour represents the effective temperature of a planet-hosting M dwarf. The four close-in, sub-Neptunes validated in this study are shown as red stars.
	\label{fig:R-e}
	}
\end{figure}

\begin{figure}[ht!]
	\epsscale{1.1}
	\plotone{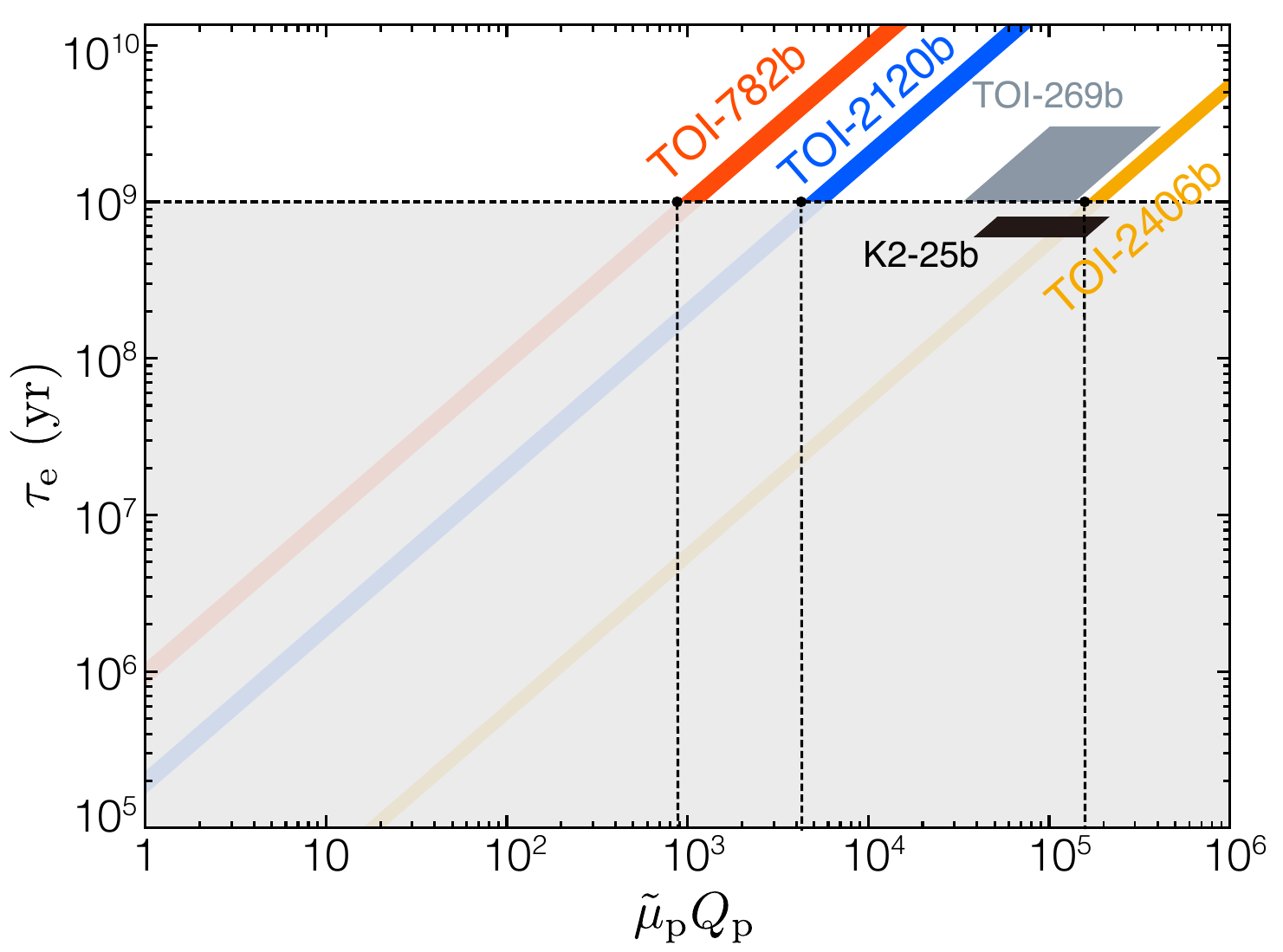}
	\caption{The eccentricity damping timescale of the three sub-Neptunes: TOI-782\,b (red), TOI-2120\,b (blue), and TOI-2406\,b (orange).
	We adopted the upper limit values of the planetary masses for TOI-782\,b, TOI-2120\,b, and TOI-2406\,b. The expected tidal dissipation factors of two other eccentric sub-Neptunes, K2-25b (black) and TOI-269\,b (gray), are also shown for comparison. The vertical dashed lines correspond to $\tilde{\mu_\mathrm{p}} Q_\mathrm{p}$ of planets with $\tau_\mathrm{e} = 1\,\mathrm{Gyr}$.
	\label{fig:tau}
	}
\end{figure}

We consider the tidal evolution of thr three short-period sub-Neptunes in eccentric orbits around relatively old M dwarfs (see Table \ref{tab:planet}). A tidal interaction between a short-period planet and an M dwarf circularizes the orbit of the planet over gigayears.
The eccentricity damping of a planet in equilibrium tide \citep[e.g.][]{1999ssd..book.....M} is written as 
\begin{equation}
	\tau_e = - \frac{e}{\dot{e}} = \frac{4}{63}\frac{M_\mathrm{p}}{M_\star}
			\left(\frac{a}{R_\mathrm{p}} \right)^5 \frac{\tilde{\mu_\mathrm{p}} Q_\mathrm{p}}{\Omega},
\end{equation}
where $e$ is the eccentricity of the planet, $M_\mathrm{p}$ is the planetary mass, $R_\mathrm{p}$ is the planetary radius, $a$ is the semimajor axis, $\tilde{\mu_\mathrm{p}}$ is the effective bulk modulus, $Q_\mathrm{p}$ is the tidal dissipation factor of the planet, $M_\star$ is the mass of the host star, and $\Omega$ is the Keplerian angular velocity. Figure \ref{fig:tau} shows the eccentricity damping timescales of the three sub-Neptunes. We adopt the upper limit masses of TOI-782\,b, TOI-2120\,b, and 2406\,b. A nonzero eccentricity for each planet can persist over gigayears if the $Q$ value is $\gtrsim 10^3$ for TOI-782\,b and 2120\,b, and $\gtrsim 10^5$ for TOI-2406\,b, assuming that $\tilde{\mu_\mathrm{p}} \sim 1$ and $Q_\mathrm{p}$ is constant for 1\,Gyr. A typical $Q$ value for a terrestrial planet and a gas giant is $10-100$ and $10^5-10^6$, respectively.
We consider here that massive rocky planets may have a $Q$-value greater than $10-100$. Massive rocky planets have magma oceans for a long period of time due to high pressures and temperatures in their interiors. Tidal damping in rocky super-Earths can be less efficient due to the size effect \citep{2012ApJ...746..150E}. Tidal dissipation in massive rocky planets may be weaker than that in terrestrial planets. On the other hand, the observed eccentricity of super-Earths suggests that their typical $Q$ value is $\sim 10-100$ \citep{2015MNRAS.448.1044H}, for example, $Q_\mathrm{p} \sim 100$ for CoRoT-7\,b \citep{2015A&A...584A..60C}. The interiors of these three eccentric sub-Neptunes with large $Q$ values are different from those of close-in super-Earths like CoRoT-7\,b.
In fact, the mass-radius relationship of these four sub-Neptunes supports that they are not composed of pure rock (see Figure \ref{fig:MR}). The four validated planets, as well as K2-18\,b \citep[e.g.][]{2023ApJ...956L..13M} and TOI-270\,d \citep{2024A&A...683L...2H,2024arXiv240303325B}, may be interesting targets for atmospheric characterization of sub-Neptunes.
Recently, \citet{2022Sci...377.1211L} showed that planets with radii larger than $1.5R_\oplus$ around M dwarfs have either water-rich interiors or significant atmospheres. These interior models are consistent with planet formation models in the course of orbital migration \citep[e.g.][]{2020A&A...643L...1V,2021A&A...656A..72B}, followed by the photoevaporation scenario of planets with different bulk compositions \citep[e.g., see Figure 7 in][]{2017ApJ...847...29O}.

Other known eccentric, sub-Neptunes close to M dwarfs, K2-25\,b with $R_\mathrm{p} = 3.44\pm0.12R_\oplus$, $M_\mathrm{p} = 24.7^{+5.7}_{-5.2}\,M_\oplus$, and $e = 0.43\pm0.05$ in the Hyades star cluster (600--800Myr; \citep{2016ApJ...818...46M, 2020AJ....160..192S}) and TOI-269\,b with $R_\mathrm{p} = 2.77\pm0.12R_\oplus$, $M_\mathrm{p} = 8.8\pm1.4M_\oplus$, and $e = 0.425^{+0.082}_{-0.086}$ \citep{2021A&A...650A.145C} around an old M dwarf with an age of a few gigayears, have drawn attention to the dynamical histories of close-in planets. K2-25\,b and TOI-269\,b can remain in an elliptical orbit until the present day if $Q_\mathrm{p} \gtrsim 10^4-10^5$ (see Figure \ref{fig:tau}). These $Q$ values larger than those of terrestrial planets are consistent with low bulk densities of both planets, indicating the presence of volatile material in their interiors.

\section{Summary}

We report the discovery and follow-up of four sub-Neptunes with radii of $2-3R_\oplus$ and $P \lesssim 8$\,days orbiting M dwarfs (TOI-782, TOI-1448, TOI-2120, and TOI-2406), three of which were newly validated by ground-based follow-up observations and statistical analyses. Our RV follow-up observations with Subaru/IRD suggest that all of the four planets are less massive than $20M_\oplus$ at 2$\sigma$ significance. The orbits of at least three short-period sub-Neptunes orbiting relatively old M dwarfs with ages $\gtrsim 1$\,Gyr are not yet tidally circularized. The data suggest that these planets have eccentricities of $e \sim 0.2-0.3$, which would require inefficient tidal dissipation in their interior. The slow damping of the eccentricities due to tidal interactions requires a large $Q$ value of $10^3-10^5$ for these three sub-Neptunes. The mass-radius relationship of all four sub-Neptunes suggests that they are unlikely to be pure rocky planets, that is, they have a significant atmosphere and/or an icy mantle on the core. The validated eccentric sub-Neptunes give us an unprecedented opportunity to infer the elusive interiors and formation histories of sub-Neptunes close to a central star.

\appendix

\section{Spectral energy distributions}

\begin{figure*}[ht]
    \gridline{\fig{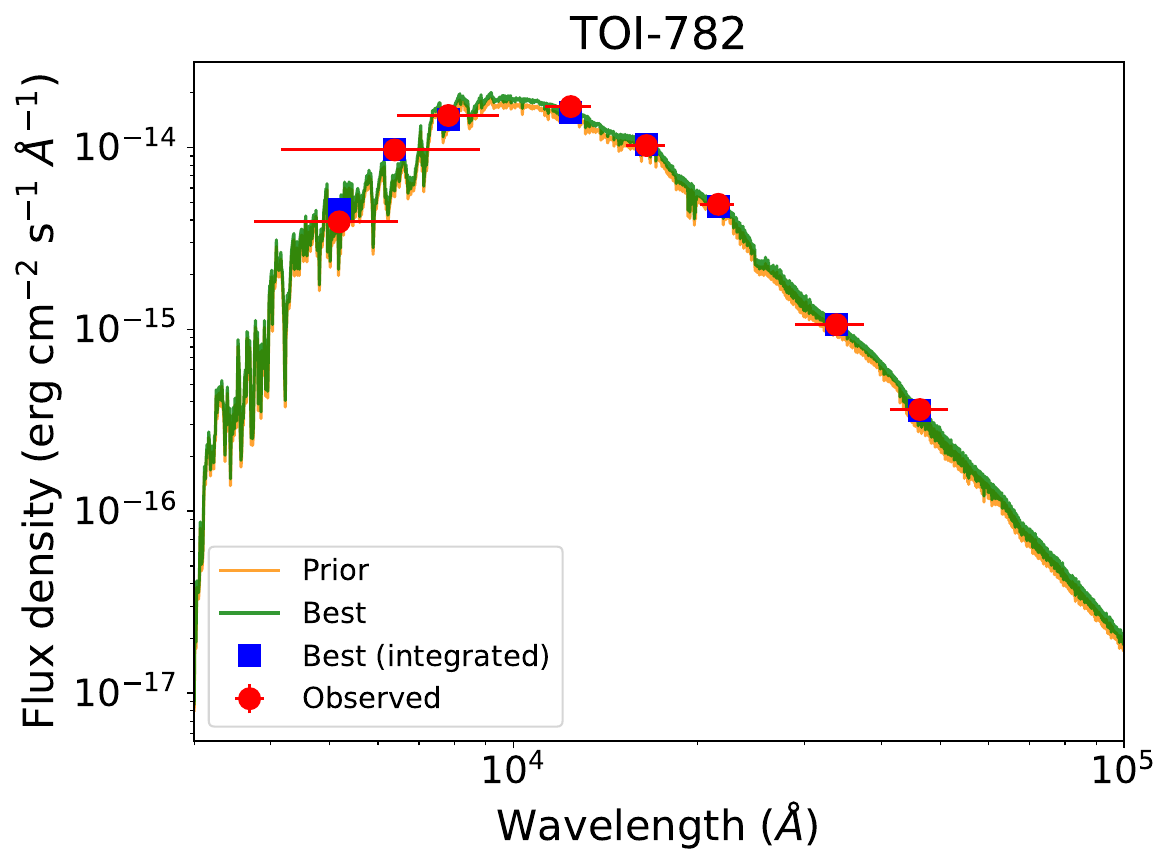}{0.4\textwidth}{(a)}
            \fig{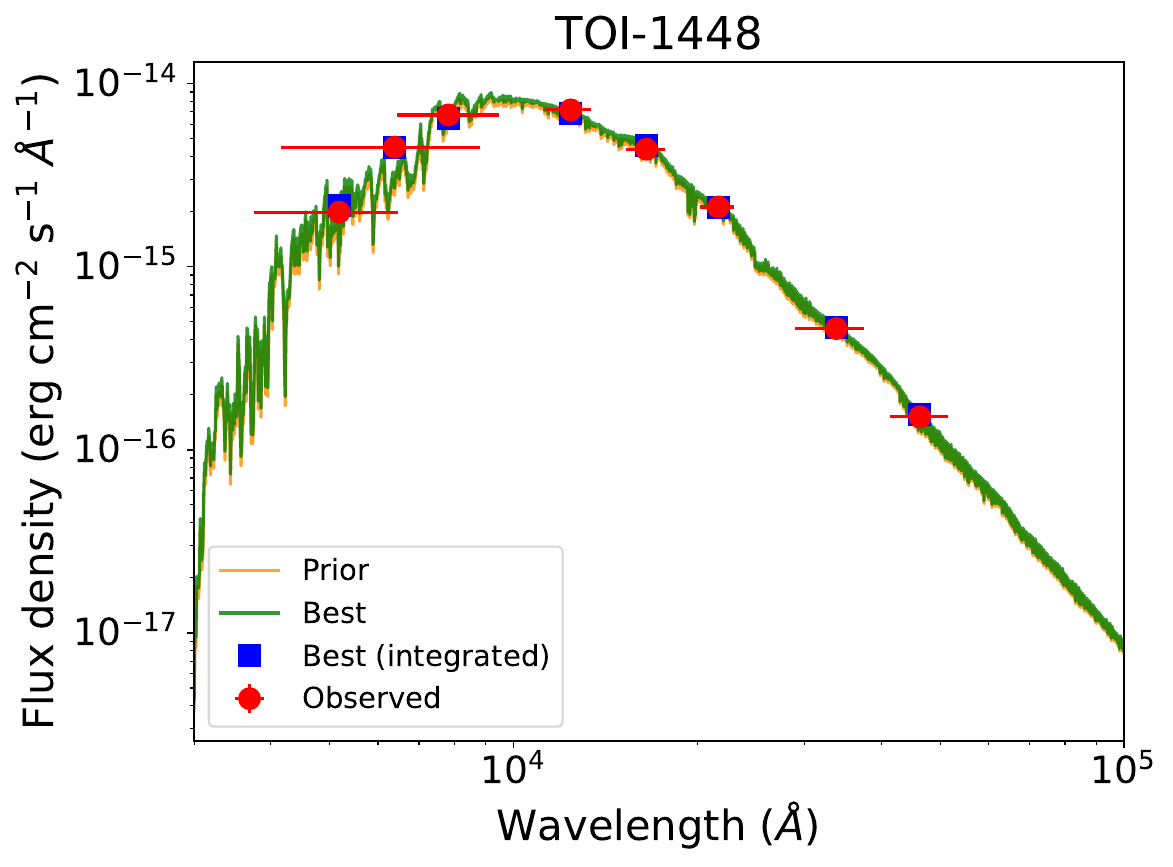}{0.4\textwidth}{(b)}}
    \gridline{\fig{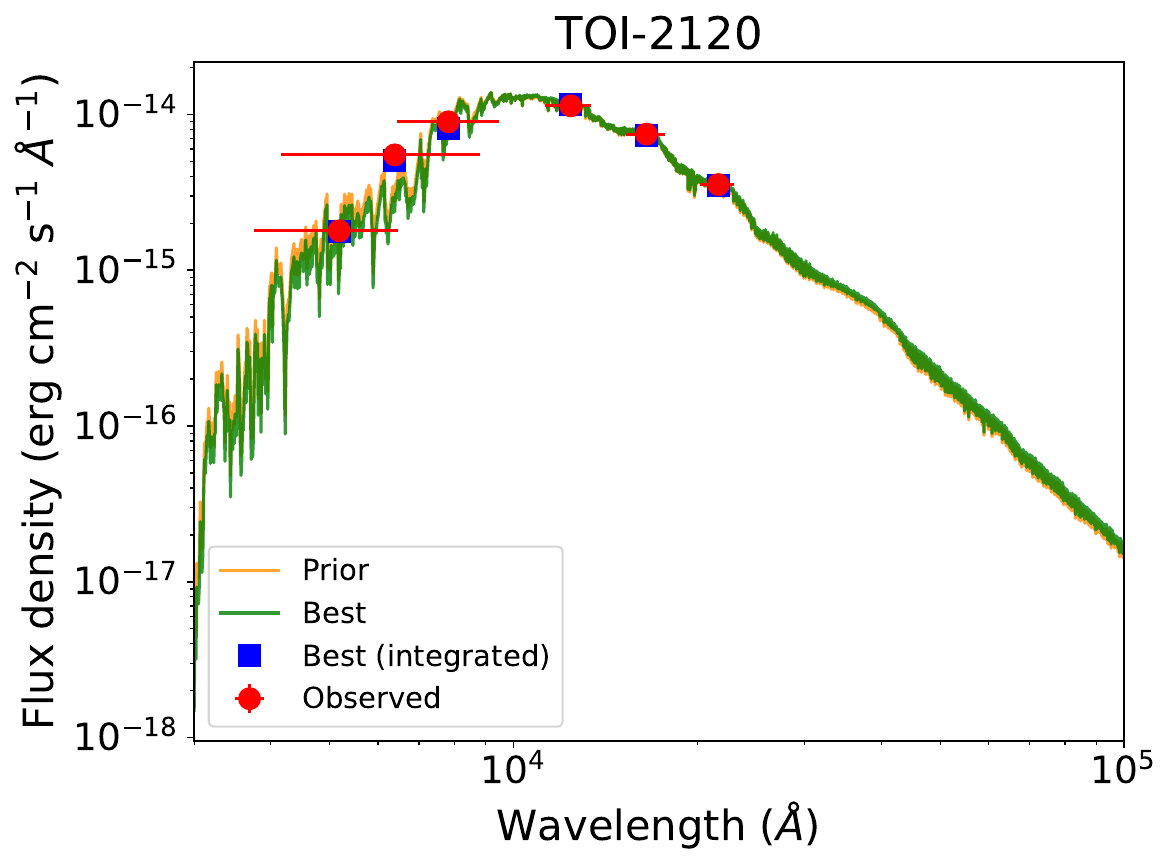}{0.4\textwidth}{(c)}
            \fig{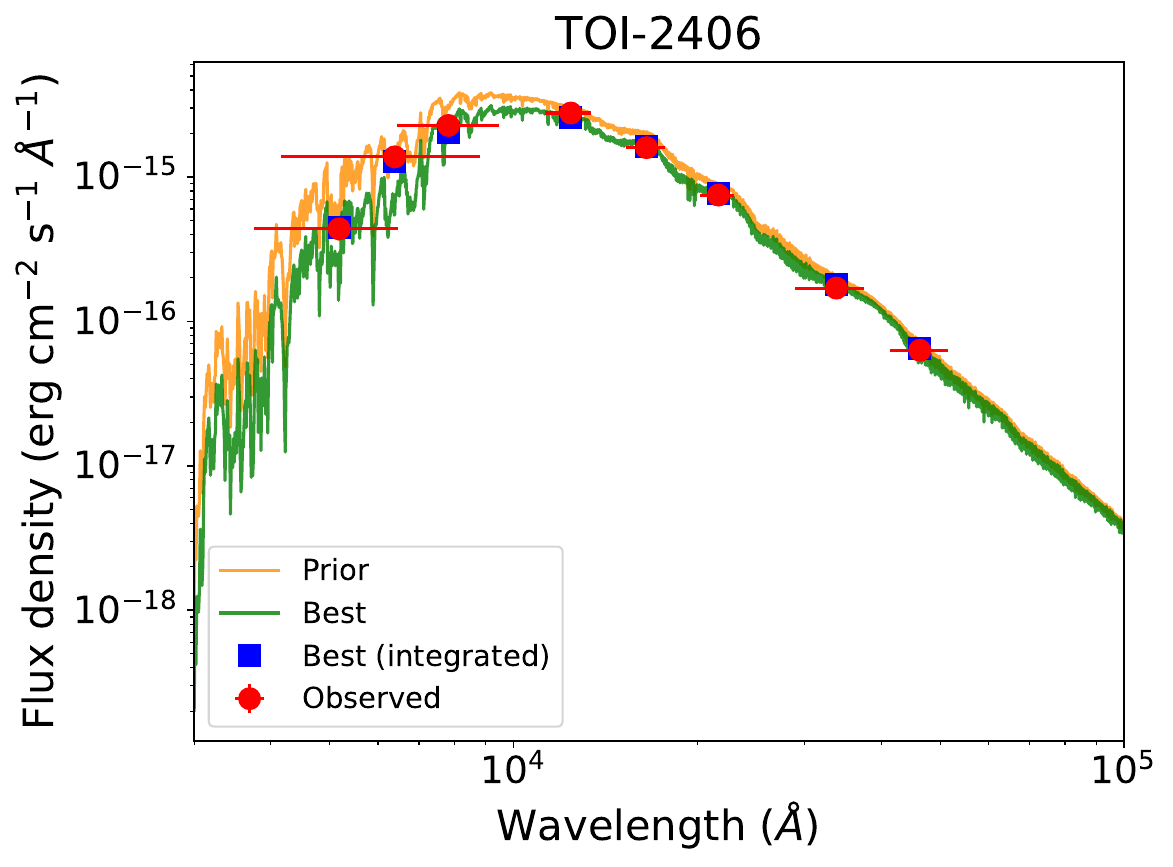}{0.4\textwidth}{(d)}}
	\caption{(a) SED of TOI-782. Panels (b), (c), and (d) are the same as (a), but for TOI-1448, TOI-2120, and TOI-2406, respectively. Yellow and blue curves show the prior and posterior SED models. Red diamonds and blue squares are observational data and best-fit results, respectively.
	\label{fig:SED}
	}
\end{figure*}

The SEDs of the four host stars are shown in Figure \ref{fig:SED}.

\section{TESS light curves of the individual sectors}

\begin{figure*}[ht]
    \centering
    \includegraphics[width=0.75\textwidth]{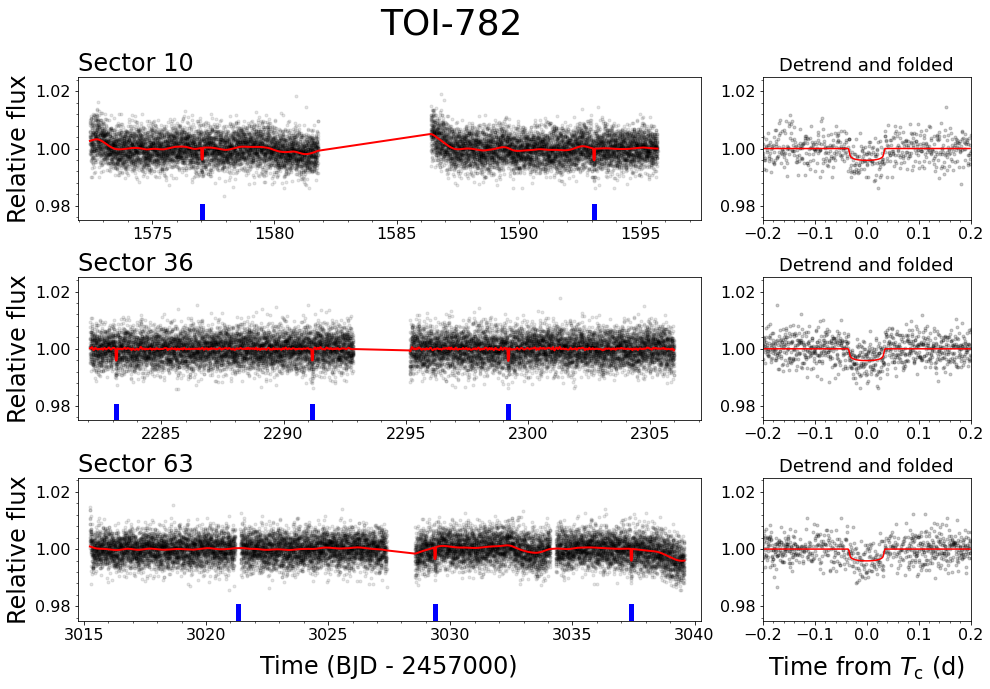}
    \caption{(Left) TESS PDCSAP light curves of TOI-782. Black dots and red lines are undetrended data points and maximum-likelihood transit+GP models, respectively. The locations of the transits covered by TESS are marked by blue vertical lines.  (Right) Same as left but the data are detrended and phase-folded by the orbital period of TOI-782b for each sector.
	\label{fig:lc_TESS_TOI-782}
	}
\end{figure*}

The TESS PDC-SAP light curves of each sector for TOI-782, TOI-1448, TOI-2120, and TOI-2406 are shown from Figures \ref{fig:lc_TESS_TOI-782} to \ref{fig:lc_TESS_TOI-782}.

\begin{figure*}[ht]
    \centering
    \includegraphics[width=0.75\textwidth]{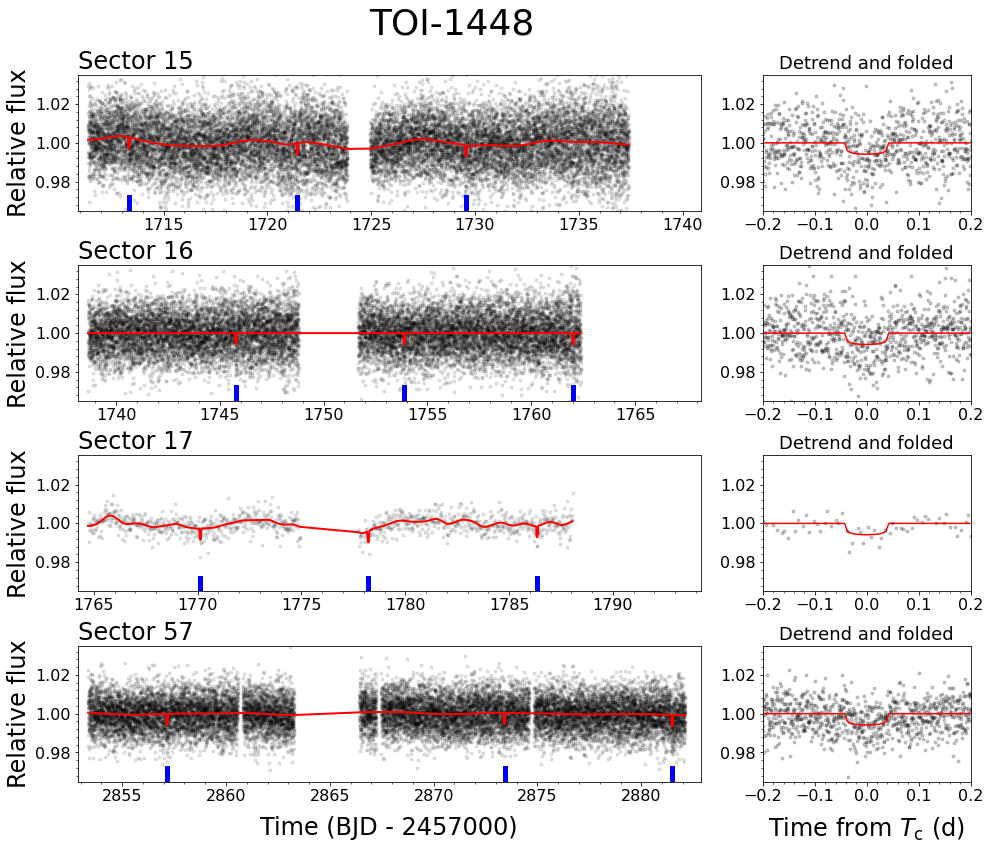}
    \caption{Same as Figure \ref{fig:lc_TESS_TOI-782}, but for TOI-1448.
	\label{fig:lc_TESS_TOI-1448}
	}
\end{figure*}

\begin{figure*}[ht]
    \centering
    \includegraphics[width=0.75\textwidth]{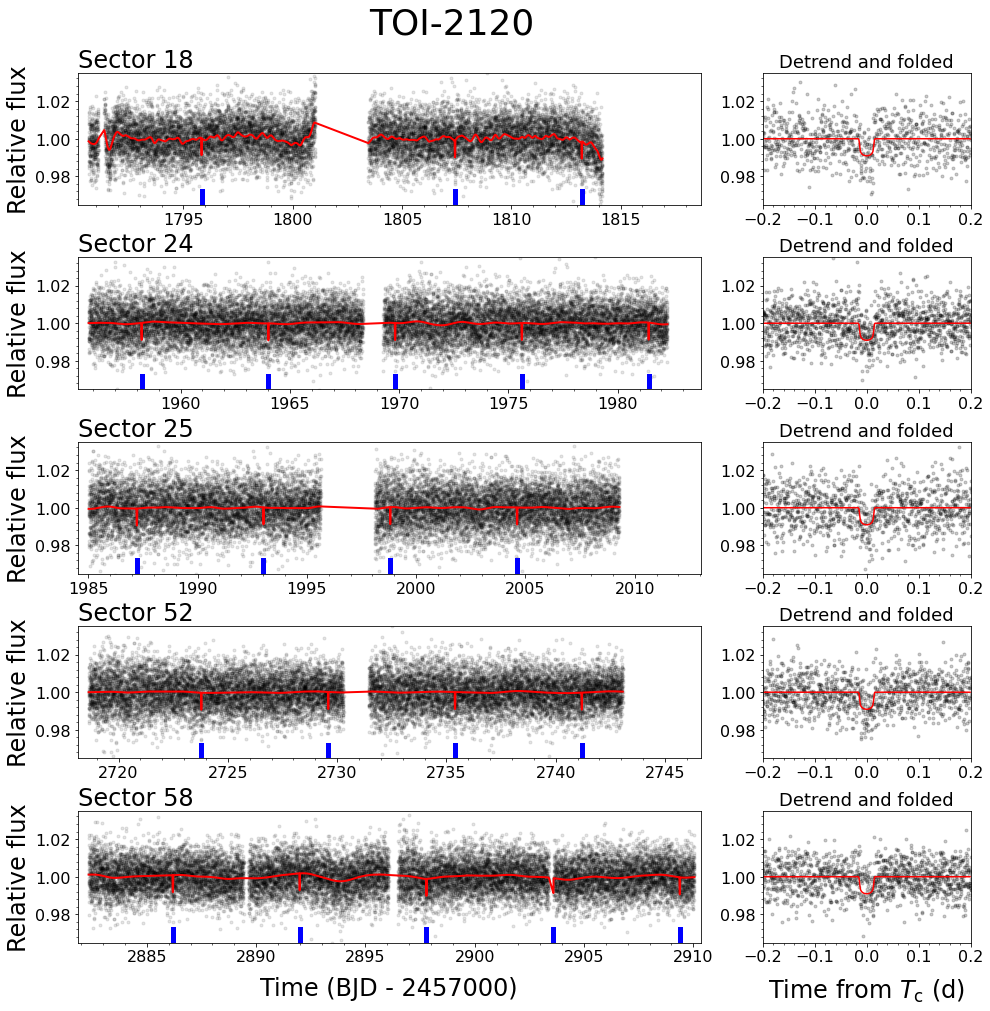}
    \caption{Same as Figure \ref{fig:lc_TESS_TOI-782}, but for TOI-2120.
	\label{fig:lc_TESS_TOI-2120}
	}
\end{figure*}

\begin{figure*}[ht]
    \centering
    \includegraphics[width=0.75\textwidth]{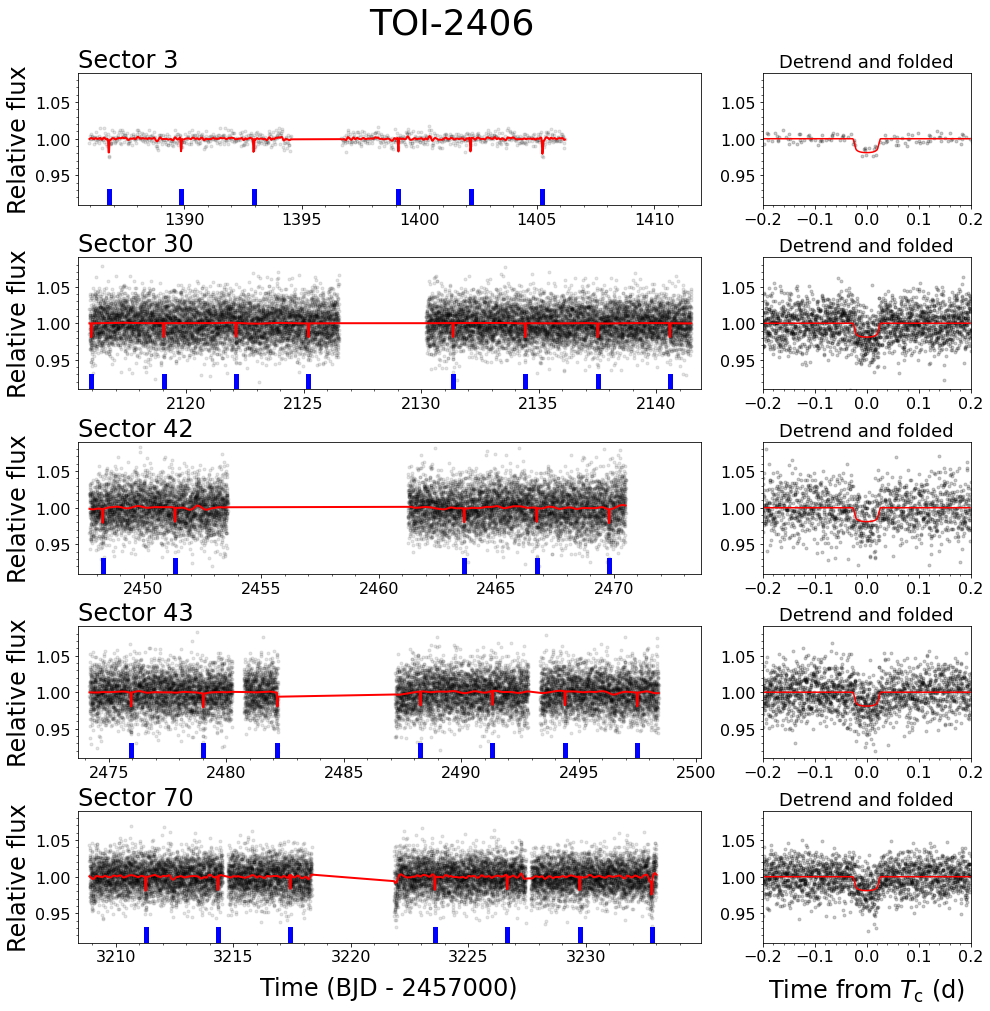}
    \caption{Same as Figure \ref{fig:lc_TESS_TOI-782}, but for TOI-2406.
	\label{fig:lc_TESS_TOI-2406}
	}
\end{figure*}

\section{Individual transit light curves from ground-based observations}

\begin{figure*}[ht]
    \gridline{\fig{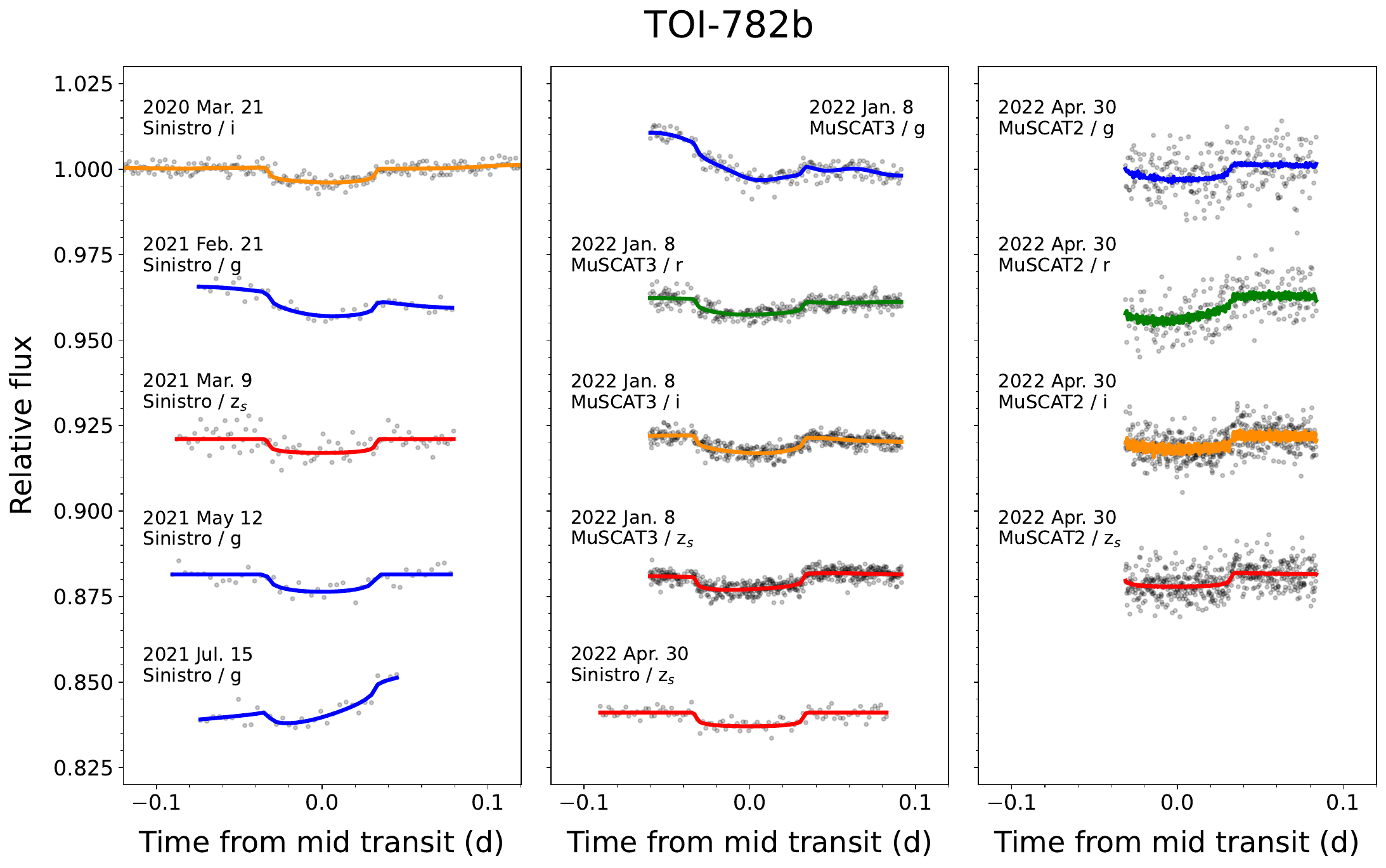}{0.55\textwidth}{(a)}
            \fig{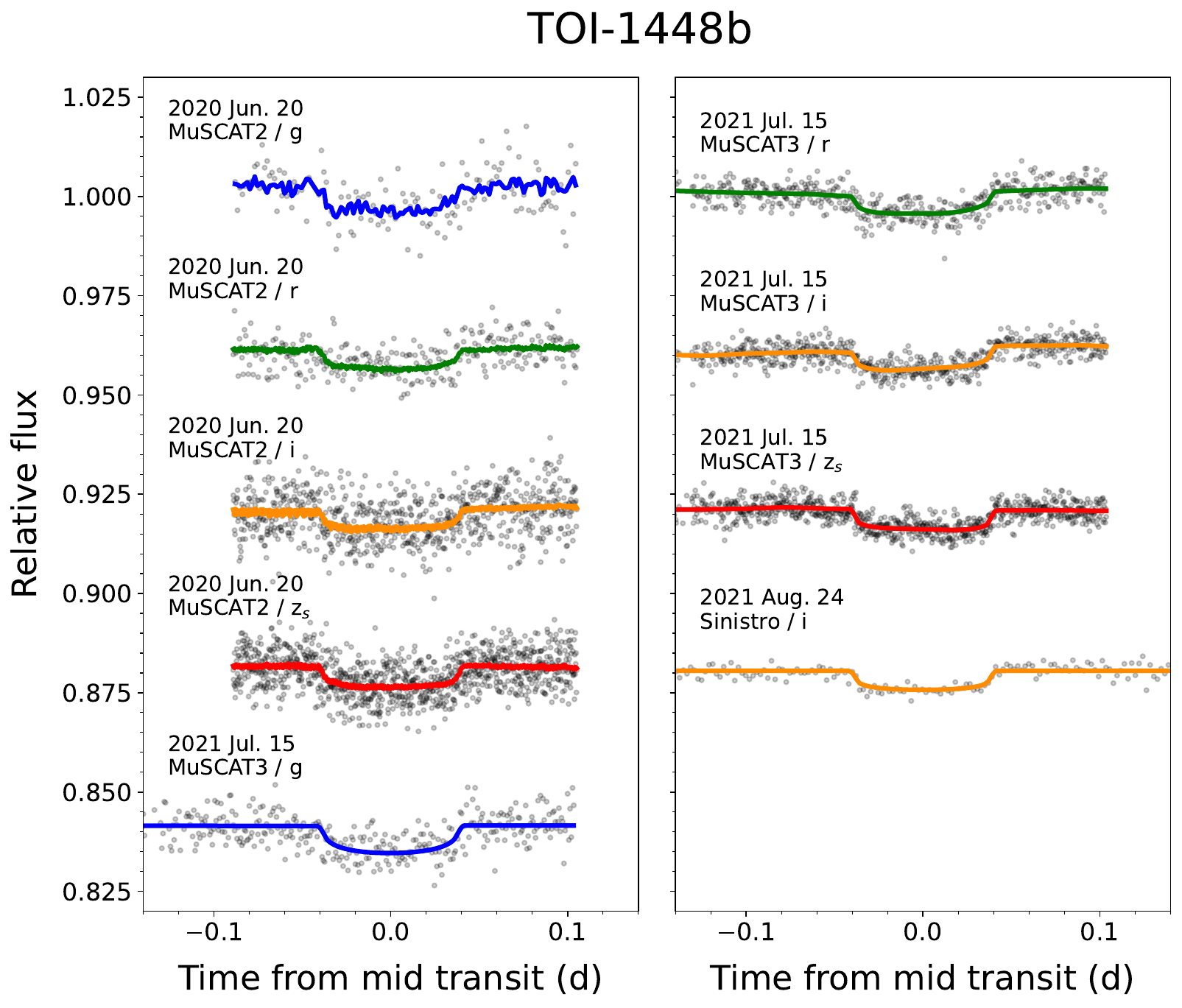}{0.41\textwidth}{(b)}}
    \gridline{\fig{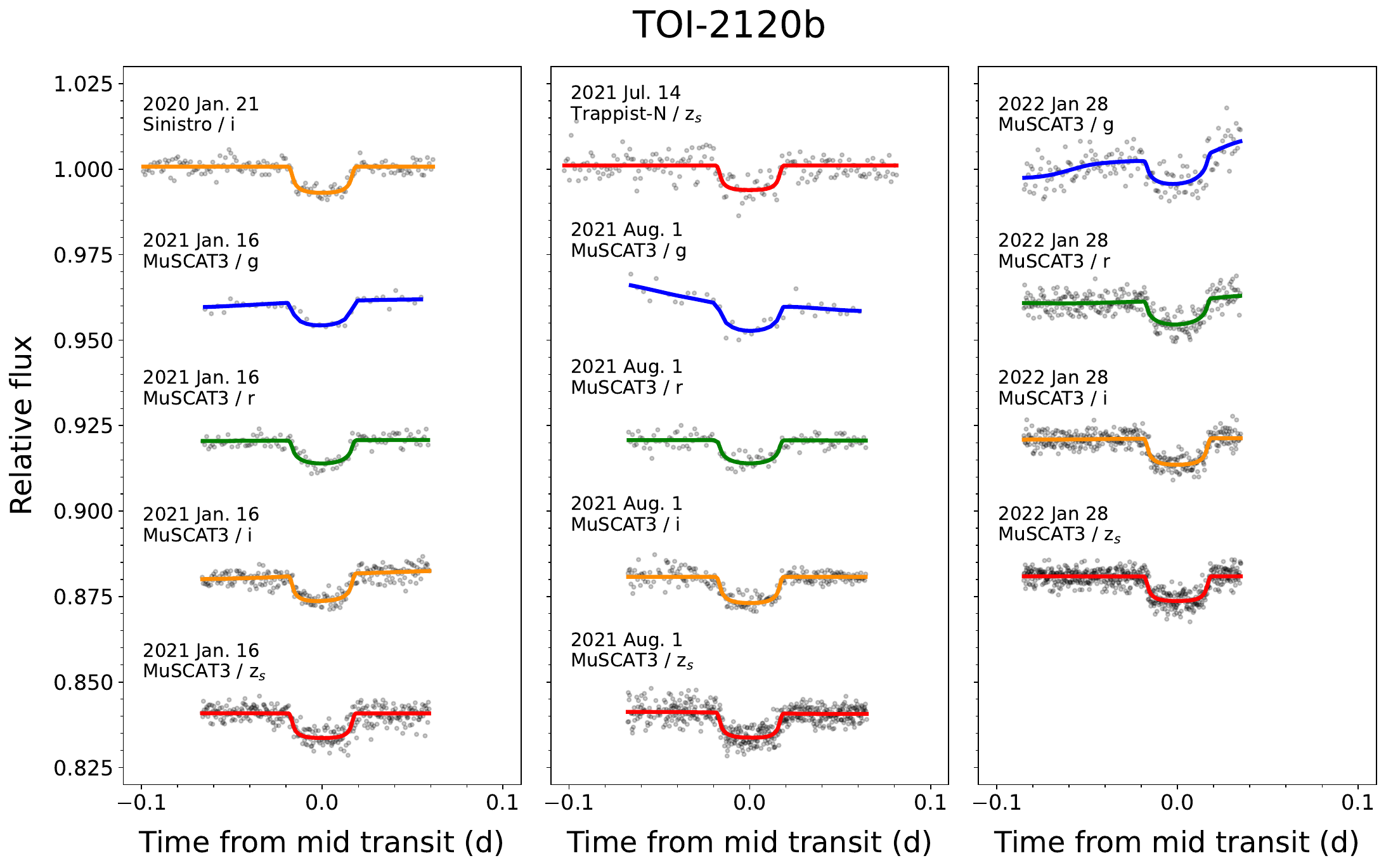}{0.55\textwidth}{(c)}
            \fig{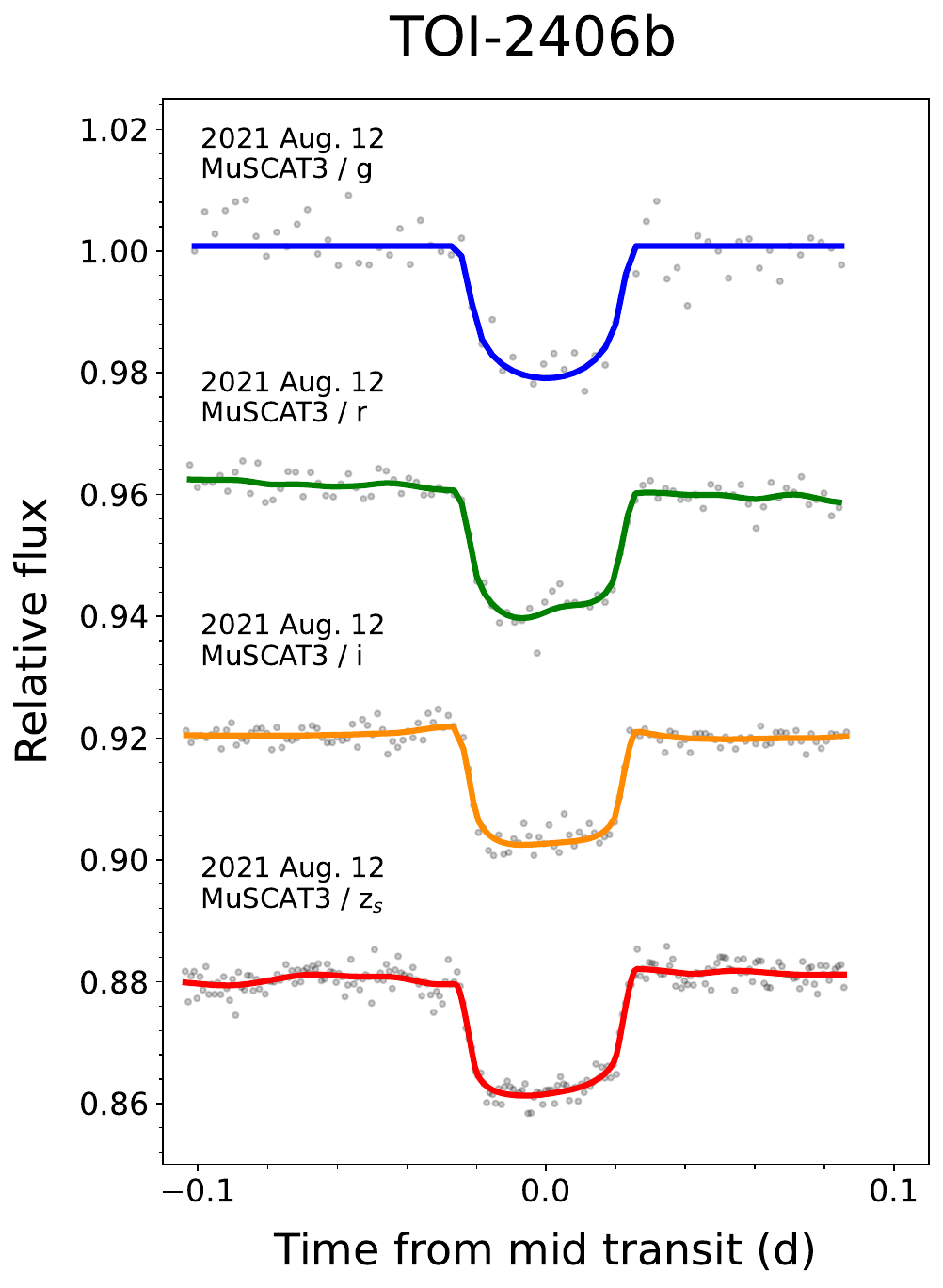}{0.25\textwidth}{(d)}}
	\caption{(a) Individual transit light curves of TOI-782\,b obtained with ground-based telescopes. Panels (b), (c), and (d) are the same as (a), but for TOI-1448\,b, TOI-2120\,b, and TOI-2406\,b, respectively.
	\label{fig:lc_ground}
	}
\end{figure*}

The individual transit light curves of the four planets obtained with the ground-based instruments are shown in Figure \ref{fig:lc_ground}.

\bibliography{reference}{}
\bibliographystyle{aasjournal}

\begin{acknowledgements}
\section*{Acknowledgments}

C.A.C acknowledges that this research was carried out at the Jet Propulsion Laboratory, California Institute of Technology, under a contract with the National Aeronautics and Space Administration (80NM0018D0004).
This work makes use of  NASA Exoplanet Archive (Confirmed Planets Table; DOI: 10.26133/NEA12) website and
the Exoplanet Follow-up Observation Program (ExoFOP; DOI: 10.26134/ExoFOP5) website, which is operated by the California Institute of Technology, under contract with the National Aeronautics and Space Administration under the Exoplanet Exploration Program and observations from the LCOGT network and the MuSCAT2-MuSCAT3 network. Part of the LCOGT telescope time was granted by NOIRLab through the Mid-Scale Innovations Program (MSIP). MSIP is funded by the NSF. The MuSCAT2 instrument at TCS operated on the island of Tenerife by the IAC in the Spanish Observatorio del Teide and the MuSCAT3 instrument at Faulkes Telescope North on Maui, HI, operated by the Las Cumbres Observatory were developed by the Astrobiology Center under financial supports by JSPS KAKENHI (18H05439) and JST PRESTO (JPMJPR1775).
K.A.C. acknowledges support from the TESS mission via subaward s3449 from MIT. Funding for the TESS mission is provided by NASA's Science Mission Directorate. D.R.C. acknowledges partial support from NASA grant 18-2XRP18\_2-0007.
This work was partly supported by a Grant-in-Aid for Scientific Research on Innovative Areas (JSPS KAKENHI Grant Number 18H05439). TRAPPIST is funded by the Belgian Fund for Scientific Research (Fond National de la Recherche Scientifique, FNRS) under the grant FRFC 2.5.594.09.F, with the participation of the Swiss National Science Foundation (SNF). I.A.S. acknowledges the support of the Ministry of Science and Higher Education of the Russian Federation under grant 075-15-2020-780 (N13.1902.21.0039). The NN-EXPLORE Exoplanet and Stellar Speckle Imager (NESSI) was funded by the NASA Exoplanet Exploration Program and the NASA Ames Research Center. NESSI was built at the Ames Research Center by Steve B. Howell, Nic Scott, Elliott P. Horch, and Emmett Quigley. Some of the observations in this paper made use of the High-Resolution Imaging instrument ‘Alopeke and were obtained under Gemini LLP Proposal Number: GN/S-2021A-LP-105. 'Alopeke was funded by the NASA Exoplanet Exploration Program and built at the NASA Ames Research Center by Steve B. Howell, Nic Scott, Elliott P. Horch, and Emmett Quigley.
'Alopeke was mounted on the Gemini North telescope of the international Gemini Observatory, a program of NSF’s OIR Lab, which is managed by the Association of Universities for Research in Astronomy (AURA) under a cooperative agreement with the National Science Foundation on behalf of the Gemini partnership: the National Science Foundation (United States), National Research Council (Canada), Agencia Nacional de Investigaci\'{o}n y Desarrollo (Chile), Ministerio de Ciencia, Tecnolog\'{i}a e Innovaci\'{o}n (Argentina), Minist\'{e}rio da Ci\^{e}ncia, Tecnologia, Inova\c{c}\~{o}es e Comunica\c{c}\~{o}es (Brazil), and the Korea Astronomy and Space Science Institute (Republic of Korea).
J.K. gratefully acknowledges the support of the Swedish National Space Agency (SNSA; DNR 2020-00104) and of the Swedish Research Council  (VR: Etableringsbidrag 2017-04945.
G.M. has received funding from the Ariel Postdoctoral Fellowship program of the Swedish National Space Agency (SNSA).
M.T. is supported by JSPS KAKENHI grant No.18H05442.
R.L. acknowledges funding from University of La Laguna through the Margarita Salas Fellowship from the Spanish Ministry of Universities ref. UNI/551/2021-May 26, and under the EU Next Generation funds.
W.W. was supported by the National Science Foundation Graduate Research Fellowship Program under grant No. DGE-1650115.
Resources supporting this work were provided by the NASA High-End Computing (HEC) Program through the NASA Advanced Supercomputing (NAS) Division at Ames Research Center for the production of the SPOC data products.
We acknowledge the use of public TESS data from pipelines at the TESS Science Office and at the TESS Science Processing Operations Center.
The research leading to these results has received funding from  the ARC grant for Concerted Research Actions, financed by the Wallonia-Brussels Federation. TRAPPIST is funded by the Belgian Fund for Scientific Research (Fond National de la Recherche Scientifique, FNRS) under the grant PDR T.0120.21. TRAPPIST-North is a project funded by the University of Liege (Belgium), in collaboration with Cadi Ayyad University of Marrakech (Morocco). M.G. is F.R.S.-FNRS Research Director and E.J. is F.R.S.-FNRS Senior Research Associate. The postdoctoral fellowship of K.B. is funded by F.R.S.-FNRS grant T.0109.20 and by the Francqui Foundation.
F.J.P. acknowledges financial support from the grant CEX2021-001131-S funded by MCIN/AEI/ 10.13039/501100011033.
This publication benefits from the support of the French Community of Belgium in the context of the FRIA Doctoral Grant awarded to M.T.
This research is based in part on data collected at the Subaru Telescope, which is operated by the National Astronomical Observatory of Japan. We are honored and grateful for the opportunity of observing the Universe from Maunakea, which has the cultural, historical, and natural significance in Hawaii.
We thank the anonymous referee for valuable comments and suggestions that improved our manuscript.
\end{acknowledgements}

\facilities{TESS, Subaru (IRD), Sanchez (MuSCAT2), FTN (MuSCAT3), SSO:1m (Sinistro), Keck:II (NIRC2), Gemini:Gillett (NIRI, 'Alopeke), SOAR (HRCam), LDT (DSSI), Hale (PHARO), WIYN (NESSI), HATSouth, Gaia, FLWO:2MASS, and WISE}

\end{document}